\documentclass[12pt,epsfig]{article}
\usepackage{epsfig}
\usepackage{color}
\newcommand{\gsim}{\lower.7ex\hbox{$
\;\stackrel{\textstyle>}{\sim}\;$}}
\newcommand{\lsim}{\lower.7ex\hbox{$
\;\stackrel{\textstyle<}{\sim}\;$}}

\def\lsim{\mathrel{\rlap{\lower3pt\hbox{\hskip0pt$\sim$}}
    \raise1pt\hbox{$<$}}}         %less than or approx. symbol
\def\gsim{\mathrel{\rlap{\lower4pt\hbox{\hskip1pt$\sim$}}
    \raise1pt\hbox{$>$}}}         %greater than or approx. symbol

\def\beq{\begin{equation}}   \def\eeq{\end{equation}}
\def\be{\begin{equation}}   \def\ee{\end{equation}}
\def\bea{\begin{eqnarray}}   \def\eea{\end{eqnarray}}

\newcommand{\bibit}[1]{\bibitem{#1}}

\newcommand{\matel}[3]{\langle #1|#2|#3\rangle}

\newcommand{\aver}[1]{\langle #1\rangle}
\renewcommand{\Im}{\mbox{Im}\,}

\newcommand{\GeV}{\,\mbox{GeV}}
\newcommand{\MeV}{\,\mbox{MeV}}

\newcommand{\La}{\overline{\Lambda}}
\newcommand{\Lam}{\Lambda_{\rm QCD}}

\begin{document}

\addtocounter{page}{-1}

\begin{flushright}
\begin{tabular}{l}
UND-HEP-01-BIG\hspace*{.2em}01\\
Bicocca-FT-01/18\\
hep-ph/0106346\\
\end{tabular}
\end{flushright}
\vspace{.3cm}
\begin{center} \LARGE \bf
{A Vademecum on
Quark-Hadron Duality}\\
\end{center}
\vspace*{.1cm}
\begin{center} {\Large
Ikaros Bigi $^{a}$\\ 
Nikolai Uraltsev $^{a,b,c}$}\\
\vspace{.8cm}
{\normalsize
$^a${\it Physics Dept.,
Univ. of Notre Dame du
Lac, Notre Dame, IN 46556, U.S.A.} }
\\
$^b$ INFN Sezione di Milano, %% Bicocca, 
University of Milano -- Bicocca, Italy\\
$^c${\it Petersburg Nuclear Physics Inst., Gatchina,
St.\,Petersburg, 188300, Russia} 
\vspace*{16mm}

{\Large{\bf  Abstract}}\\
\end{center}

We present an elementary introduction to the problem of quark-hadron
duality and its practical limitations, in particular as it concerns
local duality violation in inclusive $B$ meson decays. We show
that the accurate definition of duality violation elaborated over the
recent years allows one to derive informative constraints on
violations of local duality. The magnitude of duality violation is
particularly restricted in the total semileptonic widths. This
explains its strong suppression in concrete dynamical estimates. We
analyze the origin of the suppression factors in a model-independent
setting, including a fresh perspective on the Small Velocity
expansion. A new potentially significant mechanism for violation of
local duality in $\Gamma_{\rm sl}(B)$ is analyzed. Yet we conclude
that the amount of duality violation in $\Gamma_{\rm sl}(B)$ must be
safely below the half percent level, with realistic estimates being
actually much smaller. Violation of local duality in $\Gamma_{\rm
sl}(B)$ is thus far below the level relevant to phenomenology. We also
present a cautionary note on the $B\to D^*$ decay amplitude at zero
recoil and show that it is much more vulnerable to violations of
quark-hadron duality than $\Gamma_{\rm sl}(B)$. A critical
review of some recent literature is given. We point out that the presently
limiting factor in genuinely model-independent extraction of $V_{cb}$
is the precise value of the short-distance charm quark mass.
We suggest a direct and precise experimental check of local
quark-hadron duality in semileptonic $B\to X_c \, \ell\nu$ decays.

\thispagestyle{empty}

\vspace*{.2cm}
\vfill
\noindent
\vskip 5mm
%PACS 11.30.Er, 13.20.Eb, 13.25.Es
%\vskip 3mm

\newpage

\tableofcontents

%% \newpage

%%%%%%%%%%%%%%%%
\section{The problem}
%%%%%%%%%%%%%%

In QCD one evaluates transition rates, distributions etc. in terms of
quarks and gluons, for which the fundamental interactions are specified. 
Yet the asymptotic and thus observable states are hadrons. Some notion of
quark-hadron duality -- or duality for short -- has underlied many
applications of the quark model from the  early days on. It is  based on
the idea that a quark level calculation should at least approximate
hadronic rates. Typically it was invoked in a rather vaguely
formulated way; for example it  was not stated how accurate such an
approximation would be. A more concrete phenomenological formulation was
given in the special case of deep inelastic lepton-nucleon scattering to
extend the onset of scaling to lower momentum transfers \cite{bloom}.

Nowadays there is little doubt that QCD is the true theory
of strong interactions predicting -- in principle -- all properties of
hadrons with unlimited accuracy. The assumption of duality 
in its most general version amounts to no more than this
conviction; i.e.,  the true hadronic
observables coincide with what one ultimately obtains in the
quark-gluon language provided {\it all possible sources of
corrections} to the parton picture stemming from QCD itself are
properly accounted for. The question of duality thus has shifted to a 
different level: it amounts to assessing the potential magnitude
of those QCD contributions to the observables of interest that have 
not been included. 

The practical validity of duality thus depends on 
the theoretical tools available for treating QCD dynamics. 
The inception of QCD as a theory of strong interactions
was intimately related with realizing its asymptotically free
nature. Correspondingly, at first the quantitative treatment of strong
interactions relied almost exclusively on perturbative expansion
based on the smallness of the effective gauge coupling between quarks and
gluons in processes at small space-time intervals. It was realized
and checked in experiment that the parton ansatz improved by
calculable perturbative corrections yields a good approximation for a
wealth of high-energy Euclidean observables representing true 
short-distance physics.

Generic QCD effects for real processes in Minkowski space are quite
different and more complicated. Even at arbitrary high energies they
can acquire divergent corrections already in the perturbative
expansion, effects governed not by the small running coupling
evaluated at the scale of overall energy, etc. A number of
``infrared-stable'' \cite{sterw} Minkowskian observables were
identified which are free from these complications, thus being 
candidates for applying the concept of duality in practice. Those
included sufficiently inclusive processes which combined different
quark-gluon channels in a certain way. Likewise, quark-hadron duality
implied to be applicable if a sufficient number of hadronic
channels were included. Yet, once again, it remained somewhat
indefinite what shall constitute a sufficient number of channels. 

One of the most inclusive and {\it a priori} infrared-stable  
Minkowskian observable amenable to perturbation theory 
is the total cross section of $e^+e^-$
annihilation into hadrons.\footnote{We assume here annihilation to
light-flavor hadrons only. Production of heavy quarks has some
peculiarities making discussion of duality less transparent.} The
concept of duality there was first addressed theoretically in 
Refs.~\cite{pqw,greco} and the more specific notion of {\it local 
quark-hadron duality} was introduced.

The parton ansatz yields an energy-independent ratio
\beq
R(s)=\frac{\sigma(e^+e^- \to \mbox{hadrons})}{\sigma(e^+e^- \to
\mu^+\mu^-)}
\eeq
equal to the number of colors $N_c$ (hereafter we omit simple factors
reflecting quark charges, number of flavors, etc.). The perturbative
corrections slightly modify this:
\beq
R^{\rm pert}(s)= N_c\left( 1+\frac{\alpha_s(s)}{\pi} + \ldots \right)\;.
\eeq
Yet those are governed by the small running coupling $\alpha_s(s)$ 
and thus are not significant. Moreover, the parton cross section remains smooth
and monotonic in perturbation theory. 

The experimental cross section,
however, exhibits manifest resonance structure up to relatively high
energies. In this domain duality between the QCD-inferred cross
section and the  observed one at a given fixed energy looked problematic.
It was suggested that the equality between the two is restored if
averaging over an energy interval -- or `smearing' -- is applied.
The problem of how to compare the QCD-based and the actual hadronic
probabilities point-to-point in energy (or other kinematic variables) is
referred to  as {\em local quark-hadron duality}. 

Resonance physics, at least in the light quark sector is tightly
related to confining properties of QCD. Since the latter has nonperturbative
origin, analyzing dynamic aspects of local duality requires
control over nonperturbative effects in a consistent QCD-based
framework. 

No real progress beyond the qualitative stage outlined in
Refs.~\cite{pqw,greco} occurred for a long time. 
In most hard processes the problem could be evaded by going to
higher energy scales where duality violations are greatly
reduced. In beauty decays this option does not exist; yet data were of
less than sterling quality and therefore did not  create pressure for a
more precise theoretical treatment. On the
theoretical side there was an unsurpassed 
road block: as long as  one has very limited control over
nonperturbative effects,  there is little meaningful
that can be said about duality violations.

The general ideas of the Operator Product Expansion (OPE) have 
been applied
to quantify 
nonperturbative effects in a number of important QCD processes since
the early 1980's \cite{banda}.
Subsequently data on beauty decays became quite precise as well. It was 
also
realized that heavy flavor physics had great potential for revealing
the  presence of New Physics indirectly; such a goal places a premium
on accuracy. At the same time heavy quark theory was developed allowing
to deal with many nonperturbative aspects in heavy flavor decays in a
well-defined way. Those developments refocussed attention on the need
to deal with duality and its limitations in a 
quantitative way.

While we have no complete theory of duality violation yet, 
significant progress towards that goal has been made, specifically for
those Minkowskian observables that are described by the OPE.
Violation of local duality there is related to the asymptotic nature of
the power expansion employed by the ``practical version'' of the OPE
\cite{shifdual,shiftasi}. Such series are not sensitive to 
contributions which asymptotically behave, say, like
$\mbox{exp}(-\mbox{const}\sqrt{Q^2}/\mu_{\rm hadr})$. While they are 
exponentially suppressed in the Euclidean domain,
they turn into pure oscillations upon continuation to
Minkowski kinematics, as in the above example with 
$Q^2\!=\!-s\!=\!-E^2_{\rm
cm}$, and remain only power-suppressed in absolute magnitude due to
various `preexponential' factors. This is the way how the general 
question
of quark-hadron duality inherent in any QCD expansion gives rise to
{\it local} duality violations specific to Minkowskian kinematics.

At the same time, the OPE imposes essential constraints on 
possible (local) duality violations, which are often missed in the
literature. As a result, many effects alleged to generate violations
of local duality, especially in the total decay rates of heavy
flavors would actually signify violations
of the OPE. Thus they have little relevance to specifics of the
Minkowskian domain, but rather contradict general principles of QCD
common to any type of power expansion both in Minkowskian and
Euclidean kinematics. 

Some clear concepts for the physical origin behind duality violations
as well as their mathematical portals have been identified in 
recent years. In particular, it was shown \cite{inst} that violations
of local duality is a very general phenomenon not directly associated
with confinement or the physics of narrow resonances. 
A review of our present theoretical understanding of local
quark-hadron duality and its violation can be found in the recent
dedicated publication by Shifman \cite{shifio}.

We have to note at this point that our terminology slightly differs
from the one adopted in Ref.~\cite{shifio} dedicated specifically to
local duality. In particular, the intrinsic limitations on
the accuracy of the `practical' OPE for truly Euclidean observables
at finite mass scales are not referred to there as related to the
general issue of quark-hadron duality. As explained in a clear way in
the review \cite{shifio}, they are typically self-manifest in the OPE
itself, and thus can be viewed as expected ``natural'' or ``usual''
uncertainty. Local duality violation for Minkowskian observables is a
phenomenon over and above this. For this reason M.\,Shifman 
does not put such an
emphasis on distinguishing the terms {\it local duality violations}
and {\it duality violation} in general, and rather often uses the both
on the parallel footing. Consequently, in the terminology consistently
adopted in Ref.~\cite{shifio} there is by definition no duality
violation in Euclidean observables, but only `natural' uncertainties
of the asymptotic power expansions. On the other hand, we refer to the
latter as limitations of generic quark-hadron duality, with local
duality violation being its very specific Minkowskian aspect. Since
the focus of all the discussions lies in local duality where our
terminologies are identical, the difference should not lead to any
confusion in the practical aspects.

Quantitative tests of local duality
violations are ultimately provided by data. Yet this is not completely 
straightforward. 
For there are practical uncertainties in OPE predictions 
due to uncertainties in the 
input parameters, which are 
intrinsically unrelated to duality. E.g., variations in the values of
the strong coupling, quark masses, the leading nonperturbative condensates
etc.\ often lead to more significant uncertainties. 
The number
of clean testgrounds for local duality is thus limited in practice.

A striking demonstration of the confidence the HEP
community has in the asymptotic validity of duality was provided
by the discussion of the heavy flavor widths of the
$Z^0$ resonance. There was an about 2\% difference in the predicted
and the
observed $Z^0$ decay width into beauty which led to a lively
debate on how significant that was {\it vis-a-vie} the experimental error.
No concern was expressed about the fact that the $Z^0$ width into beauty
hadrons was
calculated for quarks, yet measured for hadrons. Moreover, the strong
coupling $\alpha_s(M_Z)$ is routinely extracted from the perturbatively
computed hadronic $Z^0$ width with a stated theoretical uncertainty 
$0.003$ which
translates into a confidence in $\Gamma_{\rm had}(Z^0)$ of about
$10^{-3}$.

This confidence derives from the fact that duality
actually represents a very
natural concept \cite{shifio}. It is based on the picture that
OPE-treatable processes
with hadrons evolve in two steps. At first hard dynamics
proceed in the femto universe characterized by large scales
like $m_Q$ or momentum transfers $\sqrt{Q^2}$. Subsequently the quarks
(and gluons) transmogrify themselves into hadrons; since this
transformation is driven by soft dynamics, this second step is
characterized by much larger distance scales. Let us in particular 
consider the decay
of a heavy quark into light flavors. 
The typical time scale for the first step is provided
by $1/m_Q$, for the second one by $1/ \Lam$ in the restframe of
a final state quark and thus by $m_Q/\Lam^2$
in the original rest frame. When the second  step
occurs, the quarks originally present are far removed from each  other.
One then expects that the second step will determine the composition of
the final state, but not gross characteristics  like total rates, 
the directions of
energetic jets etc., since those are well established by that time.

This can be illustrated in a simple quantum mechanical example.
Consider the weak decay of a heavy quark $Q$ bound to an
antiquark $\bar q$ by a potential $V(R)$ centered on $Q$. The overall
decay width in the $1/m_Q$ expansion 
is determined by the local properties of the
potential,
namely  by $V(R)$ for $R\lsim {\cal O}(1/m_Q)$. Yet the
spectrum of the final states and other more detailed properties
depend -- even at large $m_Q$ -- on the details of the potential, 
like its behavior
at finite (or even infinite)
distances $R$.\footnote{For example, if $V(R) \to \infty$ for
$R \to \infty$, the $q^{\prime}\bar q$ spectrum in
$[Q\bar q] \to l \nu q^{\prime}\bar q$ will be discrete consisting
of a series of single narrow mesons; if on the other hand the
potential saturates for finite $R$ -- modeling the possibility for
the potential string
between $q^{\prime}$ and $\bar q$ to break producing additional quark
pairs as it happens for
real QCD -- the spectrum will be continuous with many different
hadronic configurations.} Consequently, certain effect of the 
long distance 
properties of the potential must be observed at finite $m_Q$ in the
total decay width as well -- these affect local duality at a given
$m_Q$. Applied to actual heavy quarks, the real argument
is about how good an approximation duality represents at lower scales,
in particular at the beauty mass scale, and whether it ceases to be of
a numerical value at the charm scale or not.
 
This Vademecum is organized as follows. 
After illustrating the problem of local duality with the example of
the $\tau$ width, we give our Executive Summary on local duality in
heavy flavor decays in Sect.~3. Those results are explained in subsequent
sections. After introducing the theoretical arsenal in Sect.~4, we
briefly describe two implementations of nonperturbative dynamics in
Sect.~5. This framework is applied to a dedicated discussion of the
total semileptonic width of $B$ mesons in Sect.~6 complemented with
selected comments on the literature in Sect.~7. In Sect.~8 we present
remarks on the role of duality in describing  $B\to D^*\,\ell\nu$ at
zero recoil before listing conclusions in Sect.~9 and offering final
observations in Sect.~10.

%%%%%%%%%%%%%%%
\section{The scope}
%%%%%%%%%%%%%

While the practical aspects of general
quark-hadron duality are involved in any dynamical computation one
undertakes in QCD -- even the more specific aspect of local duality 
has to be 
faced in a wide range of problems -- we limit
the discussion here mainly to weak decays of heavy flavors, with  
inclusive widths there being one of the best studied subjects in
this respect. Moreover, we focus on 
semileptonic decay widths from which $|V_{cb}|$ and $|V_{ub}|$ are 
primarily extracted and where duality
violations are most constrained. To give a sense of 
the peculiarities of
{\it local} quark-hadron duality we compare this with the  
zero-recoil $B\to D^*$ transition amplitude which is closer to a 
Euclidean-type observable. As expected on general 
grounds if $m_b \sim m_c$ were to hold $B\to D^*$ would be less 
vulnerable to duality violation than $\Gamma_{\rm sl}(B)$. Yet since 
the energy release in
$b\to c\,\ell\nu$ is much larger than the {\it charm} quark mass, the
limitations of local quark-hadron duality in $\Gamma_{\rm sl}(B)$
are suppressed compared to the intrinsic uncertainties of the power
expansion for the $B \!\to\! D^*$ amplitude. 
The hadronic
width of $\tau$ leptons is considered to illustrate the numerical
aspects of local duality in a simple setting.

%%%%%%%%%%%%%%
\subsection{The hadronic $\tau$ decay width}
%%%%%%%%%%%

The observable
\beq
R_{\tau}\equiv 
\frac{\Gamma(\tau^-\to \nu_\tau +\mbox{hadrons})}{\Gamma(\tau^-\to
\nu_\tau e^-\bar\nu_e)}
\label{28}
\eeq
is the closest analogy to the integrated -- or smeared -- cross
section of $e^+e^-$ annihilation in the interval of energy up to
$m_\tau \simeq 1.77\GeV$. More precisely, up to small electroweak
radiative corrections it is expressed in terms of the spectral
densities $\rho_V$ and $\rho_A$ in the vector and axial-vector
channels:
\beq
R_{\tau} \!=\! \int_0^{M_\tau^2} \!\frac{{\rm d}s}{M_\tau^2}
\left(1\!-\!\frac{s}{M_\tau^2}\right)^2
\!\left(1+\frac{2s}{M_\tau^2}\right)
[\rho_V(s)+\rho_A(s)] = 
\frac{I_0(M_\tau^2)}{M_\tau^2} \!-3\frac{I_2(M_\tau^2)}{M_\tau^6}+
2\frac{I_3(M_\tau^2)}{M_\tau^8}
\label{30}
\eeq
(here quark masses are neglected), where the moments $I_n$ are defined as 
\beq
I_n(M) = \int_0^{M^2} {\rm d}s\, s^n \,
[\rho_V(s)+\rho_A(s)] \;.
\label{32}
\eeq

In spite of being an averaged spectral density, $R_{\tau}$ is affected
by physics underlying local duality, which can be illustrated by the
following simple argument. Let us consider the chiral limit
$m_{u,d,s}=0$ and suppose the perturbative effects can be
neglected completely, including all the anomalous dimensions of
vacuum condensates. Eq.~(\ref{30}) would then seem to state -- without any
reference to strong dynamics --  that
$R_{\tau}$ obtained in the ``practical'' OPE is at most a fourth-order
polynomial in $1/M_\tau^2$:
\beq
R_{\tau}= A_0 + \frac{A_2}{M_\tau^4} +  \frac{A_3}{M_\tau^6} 
 +  \frac{A_4}{M_\tau^8}\;,
\label{36}
\eeq
at least above a certain mass scale $M_0$. However, since hadronic
thresholds in annihilation exist at arbitrary high energy, this
certainly cannot be true. The component of
$R_{\tau}$ not contained in Eq.(\ref{36}) represents 
a violation of local duality. 
While naively almost arbitrary, in reality it obeys severe
constraints stemming from the OPE. An explicit model was analyzed in
Ref.~\cite{d2} which manifestly exhibited these features. 

A critical look at the available experimental data on $\rho_V(s)$ and 
$\rho_A(s)$ suggests that violation of local duality {\it a priori}
could be quite sizeable. This is not surprising since nearly half of the 
decay probability is due to hadronic states with invariant
masses not exceeding $1\GeV$, a resonance rather than asymptotic
regime. An attempt is often made to extract a precise value of the
strong coupling $\alpha_s(M_{\tau})$ from
$R_{\tau}$, neglecting the potential violation of local duality. This
is hardly justified theoretically \cite{shiftau}. However, the
numerical estimates of duality violation in $R_{\tau}$ yield rather
small values, at a few percent level \cite{d2}, which may sound
surprising in view of the relatively low momentum scale characteristic
to $\tau$ decays. This is indirectly confirmed by experiment.

The OPE yields the following large-$M_\tau$ expansion of $R_{\tau}$:
\beq
\frac{R_{\tau}}{N_c} = R_0 + \Delta_{{\rm pert}}(\alpha_s) +
\frac{c_2}{M_\tau^4} +  \frac{c_3}{M_\tau^6} 
 +  \frac{c_4}{M_\tau^8} \;, 
\label{40}
\eeq
where $R_0= 1-4(m_u^2 \!+\! m_d^2 +
\sin^2{\theta_C}(m_s^2\!-\!m_d^2))/M_\tau^2 + ...$ is the parton
expression, $\Delta_{\rm pert}(\alpha_s)$ represents the perturbative
series, and the last terms emerge from the nonperturbative
condensates. The $1/M_\tau^4$ and $1/M_\tau^6$ terms appear only due
to the presence of perturbative corrections or/and contain 
powers of the light quark masses. As a result, the OPE power corrections
turn out to be strongly suppressed, at a percent level.

The perturbative factor $\Delta_{\rm pert}(\alpha_s) = 1 +\alpha_s/\pi +
... $ has been well studied theoretically. Assuming the canonical values of
the condensates, Eq.~(\ref{40}) reproduces the experimental value of
$R_{\tau}$ at $\alpha_s(M_\tau) \simeq 0.32$ which would correspond
to $\alpha_s(M_Z)\simeq 0.118$, close to the standard value
extracted from the $Z$ peak physics. Here we adopt an alternative
perspective. Varying $\alpha_s(M_\tau)$ in a generous interval
$0.25\:\mbox{to}\:0.36$ we would find for the duality-violating
component $\tilde R_{\tau}$
\beq
-0.06 < \frac{1}{N_c} \delta \tilde R_{\tau} < 0.07\; .
\label{42}
\eeq
We thus see that the violation of local duality at the $\tau$ mass
scale turns out rather small, below the $10\%$ level. 

This experimental evidence does not guarantee that 
violation of local duality is universally suppressed by such a factor 
for all types of processes; on the
contrary, it is expected to be sensitive to the details of the process 
\cite{inst,d2,varenna}. Another consideration is even more
important. As will be stated later, effects of local duality violation
must oscillate as a function of energy scale with vanishing
averages. Therefore, an accidental suppression at a fixed mass scale
just near the actual $M_\tau\simeq 1.77\GeV$ cannot be
excluded.\footnote{Available experimental data allow to limit the
magnitude of duality violation in $R_{\tau}(M)$ also at the masses $M$
somewhat below than $M_\tau$, although the limit suffers from more
experimental uncertainties.}

In spite of these caveats, the hadronic $\tau$ decay width provides a
rather direct confirmation that local duality must be a
reasonable approximation above $2\GeV$ in similar circumstances, in
accord with theoretical estimates \cite{inst,d2}.

%%%%%%%%%%%%%%
\section{Executive Summary on Local Duality in Heavy Quarks}
%%%%%%%%%%%

At first it might seem that duality between the quark and hadron
description  has little chance to
hold for total decay widths.
For the OPE yields \cite{buv}
\beq
\Gamma (H_Q) \propto m_Q^5\left( 1+ 
{\cal O}\left(\frac{1}{m_Q^2}\right) \right)
\label{44}
\eeq
-- i.e., that the width of a heavy flavor hadron is 
controlled by the fifth power of the mass of the heavy {\it quark}.  Yet the
major part of this high power comes from the available phase space
(the fact most obvious in the semileptonic decays $H_Q \to \ell
\nu\,X$) 
which is determined by hadronic masses, in particular by
$M_{H_Q}$ rather than $m_Q$.  Nevertheless it turns out  
that summation over different hadronic decay channels yields the widths
computed at the quark level with their nonperturbative corrections order
by order in
$1/m_Q$ (more accurately, in the inverse energy release). This is due to
a conspiracy between the strong interaction effects in the decay
amplitudes and the hadron masses, which can explicitly be traced in  
semileptonic decays \cite{five} where the conspiracy is enforced by 
heavy quark sum rules \cite{optical}.

As explained in the Introduction, the crucial criterion for the 
theoretical analysis is whether a reaction can be treated by the
OPE. The latter does apply to sufficiently inclusive decay widths --
be they semileptonic or nonleptonic. 
In this context
the OPE is meant to yield an expansion solely in terms of 
expectation values of {\it
local} heavy quark operators evaluated for the actual hadron $H_Q$.

The practical implementation of the OPE expresses the widths as an 
expansion in $1/m_Q$ 
(or the inverse of the energy release $E_r$) with the
coefficients shaped by short-distance physics accounted for  
perturbatively:
\beq
{\cal A}(m_Q) = \sum_{k=0}^\infty c_k \frac{(\mu_k)^k}{m_Q^k}\;, 
\qquad \qquad c_k= \sum_{l=0}^\infty a_l^{(k)}
\frac{\alpha_s}{\pi}^l\;,
\label{46}
\eeq
where ${\cal A}$ is a generic (dimensionless) 
quantity and $(\mu_k)^k$ are related to
the nonperturbative expectation values. 

With only a few terms known explicitly in these series, there are 
obvious limitations in the theoretical accuracy. This problem has
little relevance to the peculiarities of {\it local} duality, and is
not even specific to OPE-treatable observables as compared to a
generic infrared-safe quantity. Duality violations are effects {\it
over and above} this trivial reason. Even in an ideal 
scenario where all terms in both expansions of Eq.~(\ref{46}) 
were known,
the OPE series would not define the exact
result ${\cal A}(m_Q)$ completely; the power expansion of the
``practical'' OPE is only asymptotic, even in the Euclidean
situation \cite{shifdual}. It is this ambiguous component that 
behaves differently for
Euclidean and Minkowskian amplitudes.

Let us list certain features of local duality that have
been clarified by theoretical considerations over the last years:

{\bf (i)} 
The primary criterion for addressing duality violation is the
existence of the OPE for the particular observable.

{\bf (ii)}   
In general local duality cannot be exact at finite masses.
It represents an approximation the accuracy of which will increase 
with the energy scales in a way that depends on the
process in question.

{\bf (iii)}   
Effects of violation of local duality can only have the form of an
oscillating function of energy ($m_Q$, $E_r$, ...), or have to be 
exponentially suppressed. 
Duality violations {\it cannot be
blamed for a systematic excess or deficit in the decay rates.}
For example, no local duality violation can convert $m_Q$  into
$M_{H_Q}$ in the total width. 

{\bf (iv)} 
The oscillating component violating local duality may be only
power suppressed. In real QCD it nevertheless is to become
exponentially suppressed as well at large enough energy, fading out as
${\rm e}^{-(E/\tilde M)^\gamma}$ with a positive $\gamma$.  
The onset
of that regime, however, can be larger than the typical hadronic scale
$\sim 1\GeV$ -- for example, it may grow with increasing $N_c$. 
The details of the asymptotics, in particular the power of energy by which
duality violation is suppressed, depends on underlying strong dynamics
and on the concrete process. This power is rather high in {\it total
semileptonic} widths of heavy quarks.

{\bf (v)}   
The OPE equally applies to semileptonic and nonleptonic total
decay rates. Likewise, both widths are
subject to violation of local duality. The difference here is
quantitative rather than qualitative; at finite heavy quark masses 
corrections are generally larger in the nonleptonic decays. In
particular, local duality violation in nonleptonic decays can be 
boosted by the accidental near-by presence of a narrow hadronic
resonance. Similar effects are extremely suppressed for semileptonic
decays. 

{\bf (vi)}   
It is not necessary to have a 
proliferation of decay channels to reach the onset of duality, either
approximate or asymptotic. An instructive example is provided by the
so-called Small Velocity (SV) kinematics in the semileptonic 
decays \cite{sv}. A
complementary nonleptonic example was identified in the exactly
solvable 't~Hooft model \cite{d22}. 

{\bf (vii)}   
A divergence in the power expansion of ``practical'' OPE
underlying violations of local duality is related to the presence of
singularities in the quark or gluon propagators at finite (or even
infinitely large) space-time intervals.\footnote{The divergence has
nothing to do with the ultraviolet divergence of separate higher-order
condensates evaluated without an ultraviolet cutoff, but is rooted 
in 
their factorial growth with the order in the $1/m_Q$ expansion.} This is
in contrast to finite-order OPE terms which account for the singularities of
interactions (for the perturbative corrections), or for the expansion of
propagators near zero space-time intervals. 
A certain class of 
nonperturbative effects (presumably strongly suppressed) comes from
small-size instantons, which are neglected in the simplest version of the
``practical'' OPE. They are {\it not} specific to local duality and are
similar in both Minkowskian and Euclidean amplitudes. Such instantons
contribute to the Wilson coefficients in the OPE computed in the
short-distance expansion. 

{\bf (viii)}   
The presence of such finite-distance singularities is a
general, rather than exceptional feature of theories with nontrivial
self-interaction. In particular, they are not directly related to
quark-gluon confinement. Consequently, duality violation exists in
general even without confinement. The latter, however, may essentially
influence the nature of the singularities.  
\vspace*{2mm}

With the number of constraints which have to be satisfied by
quark-hadron duality, the magnitude of the effects in 
$\Gamma_{\rm sl}(b\!\to\! c)$ attributable to local duality violation 
proper can hardly reach the half percent level, and is expected even
much smaller in $\Gamma_{\rm sl}(b\!\to\! u)$. 
The uncertainties associated with the general
quark-hadron duality turn out much larger for exclusive
$B\!\to\!D^{(*)}$ amplitudes due to the charm quark being only
marginally heavy. 
\vspace*{2mm}

Some of these findings will be explained in the following sections.

%%%%%%%%%%%%%%%%%
\section{Theoretical tools}
%%%%%%%%%%%%%

The central tool in describing heavy flavor decays is the OPE 
yielding series in powers of $1/m_Q$. 
The standard  application of the OPE is expanding deeply Euclidean Green
functions at small space-time intervals where it yields well
defined asymptotic series. 
On the other hand decay amplitudes 
are shaped by processes in actual Minkowski space, 
and in general are determined by
long-distance dynamics independent of the quark mass scale.
Expansions in the heavy quark mass yield only very
limited constraints under such circumstances.
Yet {\em inclusive} 
heavy flavor transitions are controlled by short distance processes, 
albeit in Minkowski space. The Minkowskian domain of physical 
interest and the Euclidean regime where the OPE is well formulated  
are connected through {\it dispersion relations} based on the analytic
properties of the transition amplitudes governing the inclusive
decay probabilities.  

To describe generic inclusive rates one starts out by defining a
transition operator $\hat T$ through 
the time ordered product of the appropriate weak Lagrangians 
$L_w$ responsible for the selected class of weak decays
\beq
\hat T(\omega)= \int {\rm d}^4 x\; {\rm e}^{-i\omega x_0}\,  
 iT \left\{L_w(x), L_w^\dagger(0)\right\}
\label{50}
\eeq
and the corresponding forward ``scattering amplitude'' off the decaying
hadron at rest
\beq
{\cal A}(\omega)= \frac{1}{2M_{H_Q}} \matel{H_Q}{\hat T(\omega)}{H_Q}\;. 
\label{52}
\eeq
The auxiliary variable $\omega$ is defined in the complex plane and is
the counterpart of the energy in $\tau$ decays. 
In some respects it is similar to a variable mass of the heavy quark.
${\cal A}(\omega)$ has usual analytic
properties in the complex $\omega$ plane. 
Its discontinuity 
at the physical point $\omega=0$ gives the
inclusive decay rate:
\beq
\Gamma_{H_Q} = \frac{1}{2M_{H_Q}} \sum_n (2\pi)^4 \delta^4(P_n\!-\!P_{H_Q}) 
\matel{H_Q}{L_w(0)}{n} \matel{n}{L_w(0)}{H_Q}= 
2\, \Im {\cal A}(0)
\;,
\label{54}
\eeq
where $n$ generically denotes all final states. 

On the other hand, the OPE expands ${\cal A}(\omega)$ in
powers of $1/(\omega\!-\! E_{\rm r})$ with $E_{r}$ denoting 
the energy release in the weak decay.  
As mentioned above, such
expansions are well behaved at complex $\omega$ away from physical
cuts. Alternatively, ${\cal A}(\omega)$ can  
be expressed as a dispersion integral over its cuts where 
the discontinuity is given by the total probabilities of
the weak processes. This yields the power expansion for the
inclusive decay rate. 

The local duality violations emerge as specific components of the
decay widths as functions of energy, to which the dispersion integrals
are poorly sensitive. In particular, they can be due to the singularities
in the forward ``scattering'' amplitudes near the physical point
$\omega=0$,\footnote{The variable $\omega$ can be visualized as a
shift in the decaying quark mass, $m_Q \to m_Q - \omega$; this is an
approximate identification affected by $1/m_Q$ corrections. Therefore,
position of singularities at real $\omega$ determines heavy quark
masses where nonanalytic behavior occurs.} that is near the masses $m_Q$
where the decay amplitudes are singular, due to the thresholds for 
new channels or narrow resonances. 

In nonleptonic decays the operator $T(\omega)$ is most generally given by
the product of three Green functions of the final-state quarks,
contracted with $\bar{Q}(x)$ and $Q(0)$. The quark propagators are affected
by a gluon medium or by quark condensates,
Fig.~1. The singularities at physical $\omega$ can {\it a priori}\, be 
quite arbitrary and essentially depend on strong dynamics. For
example, ${\cal A}(\omega)$ can have a pole $1/(m_Q\!-\!m_0 +
\frac{i}{2}\Gamma)$ if there is an appropriate resonance at the mass
$m_0$. Such a contribution is accounted for by the OPE which, in practice 
effectively averages the width. Therefore, the OPE does not
distinguish a regular resonance structure from a smooth behavior
obtained by averaging the width over an interval of $m_Q$. As a
result, the deviation of the nonleptonic widths from the OPE
predictions at certain discrete values of $m_Q$ can, in principle,
be arbitrarily large if narrow resonances existed with large enough masses.

\thispagestyle{plain}
\begin{figure}[hhh]
\vspace*{-4mm} 
 \begin{center}
 \mbox{\epsfig{file=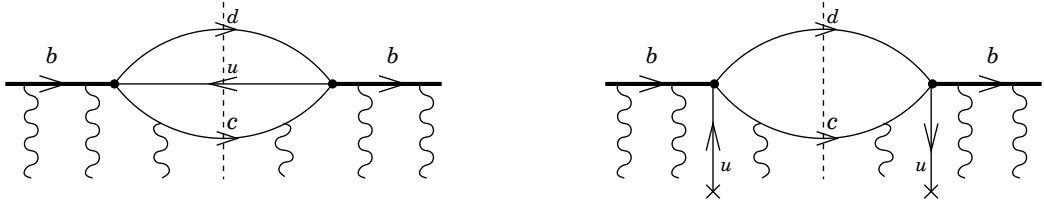,width=14cm}}
 \end{center}\vspace*{-8mm} 
\caption{ \footnotesize
Examples of the diagrams generating the transition operator
$T(\omega)$. The variable $\omega$ can be thought to enter in the weak
vertices where the amount of (complex) energy is sucked away or pumped
in. Computing the inclusive width we take the absorptive part of the 
corresponding amplitude.
}
\end{figure}

The situation is quite different in the semileptonic widths\footnote{Or
for decays like $b\to s+\gamma$, $b\to s+\ell \bar\ell$ which we do not
address here.} where two of the Green functions are replaced by 
free lepton propagators, $\frac{i}{2\pi^2} \frac{\not x}{x^4}$ 
in the coordinate
representation. This essentially limits the structure of the
singularities and strongly suppresses them at physical $\omega$. The
maximal possible violation of local duality is drastically reduced.

%%%%%%%%%%%%%%%%%%%%
\subsection{Semileptonic decays -- general formalism 
\label{SLGF}}
%%%%%%%%%%%%%%%%

In semileptonic transitions the lepton momentum is an additional 
kinematic variable. In order to study the differential distributions,
one considers a simpler 
forward scattering amplitude involving two weak quark currents:  
\beq
h_{\mu\nu} = \frac{1}{2M_{H_Q}}
\matel{H_Q}{\int {\rm d}^4 x \:{\rm e}^{-ikx}\:
iT\left\{J_\mu(x)J_\nu^\dagger(0)\right\} } {H_Q}\;,
\label{60}
\eeq
with $J_\mu=\bar{q}\gamma_\mu(1\!-\!\gamma_5)Q$ for $Q\to q$ weak
decays. From now on we assume explicitly the case of $B$ meson 
decays, and also
neglect, for simplicity, lepton masses. Decays leading to a 
$\tau\nu_\tau$ pair do not introduce essential new features. 

The tensor $h_{\mu\nu}$ is generally decomposed into five invariant
functions $h_i$ depending on $k^2$ and $k_0=kv$:
\beq
h_{\mu \nu} = - h_1 g_{\mu \nu} + h_2 v_{\mu}v_{\nu} -
i h_3\epsilon _{\mu \nu \alpha \beta} v^{\alpha}k^{\beta}
+  h_4 k_{\mu}k_{\nu} +  h_5 (k_{\mu}v_{\nu} + v_{\mu}k_{\nu})\;,
\label{62}
\eeq
with $v_\mu$ denoting the 4-velocity of the decaying meson, and their
discontinuities are the heavy quark structure functions:
\beq
w_i = 2 \, \Im h_i \;.
\label{64}
\eeq
For massless leptons $w_4$ and $w_5$ do not enter; moreover, $w_3$
does not affect the decay width:
$$
\Gamma_{\rm sl} \!=\! \frac{G_F^2 |V_{qb}|^2}{16\pi^4}
\int_{k^2>0} \!{\rm d}k^2   \int_{k_0>\sqrt{k^2}} {\rm d}k_0
\sqrt{k_0^2\!-\!k^2} \left(k^2 w_1(k_0;k^2) 
+\frac{1}{3}(k_0^2\!-\!k^2) w_2(k_0;k^2) \right) =
$$
\beq
\frac{G_F^2 |V_{qb}|^2}{16\pi^4}
\int_{0}^{\infty} \!{\rm d}\vec{k}^{\,2}\, |\vec{k}\,|   
\int_{k_0>|\vec{k}\,|} {\rm d}k_0
 \left( (k_0^2\!-\!\vec{k}^{\,2}) w_1(k_0;k_0^2\!-\!\vec{k}^{\,2}) 
+\frac{1}{3}\vec{k}^{\,2}\, w_2 (k_0;k_0^2\!-\!\vec{k}^{\,2})\right)
\label{66}
\eeq
(the upper limits of integrations are fixed by dynamics rather than
by kinematic constraints; namely, $w_i$ vanish for
$k_0>(M_B^2\!+\!M_D^2\!-\! k^2)/2M_B$, and likewise for $b\to u$.)
This expression shows that the singularities in the semileptonic
width are suppressed compared to what one finds in the general case for 
nonleptonic decays. The contribution of the particular hadronic state
$X$ with the invariant mass $M_X$ to the structure functions $w(k_0;
k^2)$ has the form 
\beq
\delta W(k_0;k^2) \propto |F_X(k^2)|^2 \,
\delta\left(k_0-\frac{M_B^2-M_X^2+k^2}{2M_B}\right)\;,
\label{68}
\eeq
where $F_X$ denote the $B\to X$ transition amplitudes up to certain
kinematic factors. Since $F_X(k^2)$ are analytic at physical $k^2$
accessible in the decays, the integration in Eqs.~(\ref{66}) provides
an amount of smearing suppressing the threshold singularities. 

Representation (\ref{66}) itself is not, however, most convenient to
reveal the related suppression, since the interval of integration over
$k_0$ shrinks to zero as $k^2$ approaches 
$E^2_{r}=(m_b\!-\!m_c)^2$. 
The duality properties of the total semileptonic width are more
explicit if it is represented in the following form:
$$
\Gamma_{\rm sl} = \frac{G_F^2 |V_{qb}|^2}{16\pi^4}
\int_{0}^\infty {\rm d}k_0  \; \int_{0}^{k_0^2} {\rm d}k^2
\sqrt{k_0^2\!-\!k^2} \left(k^2 w_1(k_0;k^2) 
+\frac{1}{3}(k_0^2\!-\!k^2) w_2(k_0;k^2) \right) =
$$
\beq
\frac{G_F^2 |V_{qb}|^2}{16\pi^4}
\int_{0}^\infty {\rm d}k_0  \;
\int_{0}^{k_0^2} {\rm d}\vec{k}^{\,2} |\vec{k}\,| 
 \left( (k_0^2\!-\!\vec{k}^{\,2}) w_1(k_0;k_0^2\!-\!\vec{k}^{\,2}) 
+\frac{1}{3}\vec{k}^{\,2}\, w_2 (k_0;k_0^2\!-\!\vec{k}^{\,2})\right)\,,
\label{72}
\eeq
or
$$
\Gamma_{\rm sl} = \frac{G_F^2 |V_{qb}|^2}{16\pi^4}
\int_{0}^\infty {\rm d}k_0  \; u(k_0)\;,
$$
$$
u(k_0)=
\int_{0}^{k_0^2} {\rm d}k^2
\sqrt{k_0^2\!-\!k^2} \left(k^2 w_1(k_0;k^2) 
+\frac{1}{3}(k_0^2\!-\!k^2) w_2(k_0;k^2) \right) \; = \qquad\qquad
$$
\beq
\int_{0}^{k_0^2} {\rm d}\vec{k}^{\,2} |\vec{k}\,| 
 \left( (k_0^2\!-\!\vec{k}^{\,2}) w_1(k_0;k_0^2\!-\!\vec{k}^{\,2}) 
+\frac{1}{3}\vec{k}^{\,2}\, w_2 (k_0;k_0^2\!-\!\vec{k}^{\,2})\right)\,.
\label{74}
\eeq
As in the standard representation (\ref{66}), the hadronic function 
$u(k_0)$ actually
vanishes at $k_0>M_B\!-\!M_D$. Likewise it can be represented
as a discontinuity of a function $\tilde W(k_0)$ related to the transition
amplitude:
\beq
\tilde W(k_0)=
\int_{0}^{k_0^2} {\rm d}k^2
\sqrt{k_0^2\!-\!k^2} \left(k^2 h_1(k_0;k^2) 
+\frac{1}{3}(k_0^2\!-\!k^2) h_2(k_0;k^2) \right) \;,
\label{76}
\eeq
$$
u(k_0)= \frac{1}{i} \, \left[\tilde W_-(k_0)\!-\!\tilde W_+(k_0)\right]\;,
$$
where $\tilde W_{\pm}(k_0)$ are obtained by integrating over $k_0$
above and below the real axis, respectively. $\tilde W(k_0)$ has
analytic properties similar to the transition amplitude $h_{\mu\nu}$
at fixed $k^2$ or $\vec{k}^{\,2}$. In particular, since the
integration in Eq.~(\ref{76}) has fixed end points, $\tilde W(k_0)$ acquires
singularities at $k_0$ where $h(k_0,0)$ or $h(k_0,k_0^2)$ become 
singular. The
former corresponds to the thresholds with vanishing lepton invariant
mass $k^2$: $k_0=\frac{M_B^2-M_X^2}{2M_B}$. The latter corresponds to
the zero-recoil thresholds $k_0=M_B\!-\!M_X$. It actually has a
special explicit counterpart in the OPE since it represents the
maximal-$k^2$ part of the decay probability \cite{WA}. However, due to
the integration over $k_0$ in the first of Eqs.~(\ref{74}), 
this singularity is not
important, and we do not go into further details here. 

The total width is obtained as an integral over the discontinuity $u(k_0)$
of $\tilde W(k_0)$ along the cut down to $k_0=0$, and therefore can be
represented as a contour integral 
\beq
\Gamma_{\rm sl} = \frac{G_F^2 |V_{qb}|^2}{16\pi^4} \frac{1}{i}
\int_{C} {\rm d}k_0  \;\tilde W(k_0) \;,
\label{77}
\eeq
see Fig.~2. This integral is not an analytic function of energy
release or of the heavy quark masses since the final points of the
contour are fixed at $k_0=0\pm i0$ lying on the cut. However, this
shows that the singularities of the width and, correspondingly the
violation of local duality is determined by the transitions with
$k^2\!=\!0$ (say, the thresholds with $M_X\!=\!M_B$) in spite of the
kinematics with $k^2\!>\!0$ contributing to the total rate. (A similar
fact was explicitly observed for instanton effects in
Ref.~\cite{inst}.) The full hadronic energy release $m_b\!-\!m_c$
controls the total width.

\thispagestyle{plain}
\begin{figure}[hhh]
\vspace*{-2mm}  
 \begin{center}
 \mbox{\epsfig{file=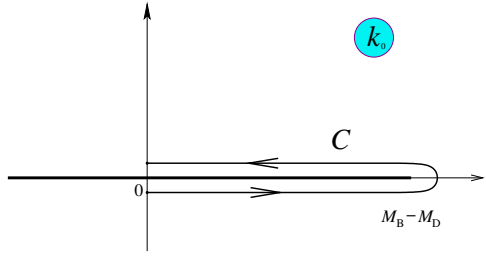,width=6.5cm}}
 \end{center}\vspace*{-8.5mm} 
\caption{ \footnotesize
The integration contour around the cut of $\tilde W(k_0)$ 
for the total semileptonic width. The end
points at $k_0=0$ are shifted into the complex plane for clarity.
}
\end{figure}

As before, this can also be shown by  
analytically continuing the correlator associated with the total
width, using the auxiliary variable $\omega$. The
amplitude ${\cal A}(\omega)$ of Eq.~(\ref{52}) is literally given 
by 
$$
{\cal A}(\omega) = 
-\frac{G_F^2 |V_{qb}|^2}{32\pi^5i} 
\int {\rm d}k_0 \,{\rm d}{\vec k}^{\,2} \,|{\vec k}\,| 
\ln{({\vec k}^{\,2}\!-\!k_0^2)}
\left((k_0^2\!-\!{\vec k}^{\,2}) h_1(k_0\!+\!\omega;k_0^2\!-\!{\vec
k}^{\,2})\: + \qquad \qquad\right.
$$
\vspace*{-4mm}
\beq
\left.
\qquad \qquad \qquad \qquad \qquad \qquad  
\frac{1}{3} {\vec k}^{\,2} 
h_2(k_0\!+\!\omega;k_0^2\!-\!{\vec k}^{\,2}) \right) \, .
\label{78a}
\eeq
For our purposes it  can instead be defined directly as
$$
{\cal A}(\omega) \!=\! 
\frac{\!G_F^2 |V_{qb}|^2}{16\pi^4\!} 
\int_{0}^M \!\!{\rm d}k_0\!
\int_{0}^{k_0^2} \! {\rm d}k^2  \sqrt{k_0^2\!-\!k^2} 
\left(k^2 h_1(k_0\!+\!\omega;k^2) \!+\!
\frac{1}{3}(k_0^2\!-\!k^2) h_2(k_0\!+\!\omega;k^2) \right) ,
$$
or
$$
{\cal A}(\omega) = 
\frac{G_F^2 |V_{qb}|^2}{16\pi^4} 
\int_{\omega}^M {\rm d}k_0
\int_{0}^{(k_0\!-\!\omega)^2} {\rm d}k^2  \sqrt{(k_0\!-\!\omega)^2\!-\!k^2} 
\left(k^2 h_1(k_0;k^2)\;  + \qquad \qquad \qquad 
\right.
$$
\vspace*{-4mm}
\nopagebreak 
\beq
\left.
\qquad \qquad\qquad \qquad\qquad \qquad \qquad 
\frac{1}{3}((k_0\!-\!\omega)^2\!-\!k^2) h_2(k_0;k^2) \right)
\label{78b}
\eeq
($M$ is a sufficiently large constant), so that for real $\omega$
\beq
2\,\Im {\cal A}(\omega) = 
\frac{G_F^2 |V_{qb}|^2}{16\pi^4} 
\int_{\omega}^{\infty} {\rm d}k_0  \;\tilde u(k_0;\omega) 
\label{79}
\eeq
with
\beq
\tilde u(k_0;\omega)\!=\!  
\int_{0}^{(k_0\!-\!\omega)^2} \! {\rm d}k^2  \sqrt{(k_0\!-\!\omega)^2\!-\!k^2} 
\left(k^2 w_1(k_0;k^2) +
\frac{1}{3}((k_0\!-\!\omega)^2\!-\!k^2) w_2(k_0;k^2) \right) . 
\label{79a}
\eeq
The approximate scaling holds 
${\cal A}(\omega; m_b)\simeq  {\cal A}(0;m_b\!-\!\omega;)$,
and one obviously has $\tilde u(k_0;0)= u(k_0)$.

%%%%%%%%%%%%
\subsection{Sum rules}
%%%%%%%%%%%%%%

Sum rules for semileptonic transition amplitudes refer to a somewhat
different type of inclusive probabilities sharing some 
peculiarities of total semileptonic widths.  
Besides providing a wealth of
important dynamic constraints, they help to illustrate and visualize a
number of the OPE results obtained directly for the total widths.
In particular, they allowed to show in the most
general setting that the total decay probability coincides with that of
the free quark ansatz, and how the hadron masses get replaced by the quark
masses thus extending this equality to the level of $1/m_Q$ corrections.

In sum rules we fix one of the two kinematic variables,
usually $k^2$ or $\vec{k}^{\,2}$, and then consider the total
transition probability weighted with a power of energy $k_0$, say
\beq
I_n^{(i)}(\vec{k}^{\,2}) = \frac{1}{2\pi} \,\int {\rm d}k_0\: 
(-\!k_0)^n\, w_i(k_0;
k_0^2\!-\!\vec{k}^{\,2}) 
\label{70}
\eeq
(in the heavy quark expansion we actually count energy from a
different value rather than $k_0=0$: we put
$\epsilon=m_b\!-\!\sqrt{m_c^2+\vec{k}^{\,2}} - k_0$ 
so that $\epsilon\!=\!0$ 
corresponds to the free quark kinematics).
Dispersion relations directly equate these moments to the coefficients
of the asymptotic expansion of the hadronic functions $h_i$ in powers
of $1/k_0$ at large (complex) $k_0$ (or $\epsilon$) computed in the OPE. 
The moments considered in the sum rules are similar to inclusive
differential decay widths at fixed $k^2$ (or
$\vec{k}^{\,2}$). However, they are
not restricted by the kinematic constraints and include states with
high mass which may not be accessible in the actual decay, but only in
the scattering of the weak current on the heavy hadron. For this
reason, the sum rules are in a sense exact relations not affected by
local duality violations. Yet physics of duality is present in the
sum rules as a question at which mass scale and how accurately the
integrals are saturated and where their normalization point dependence
enters the perturbative stage. This scale is directly related to the
onset of local duality in the actual total semileptonic widths. 

Another aspect of duality in the sum rules, again not directly related to 
local duality, is of how accurate is the expansion in the inverse charm
mass. We can compute the moments in QCD expanding the transition
operator in powers of $\Lam/E_q$ with
$E_q=\sqrt{m_q^2+\vec{k}^{\,2}}$; in practice one typically has $E_q\approx
m_c$. Here one encounters a new aspect called
{\it global duality} \cite{optical}. It was shown that the
quark-hadron correspondence holds separately for the decay-type charm
intermediate states and those which
are due to other physical processes in the full transition
amplitude $h_{\mu\nu}$, order by order in the $1/E_q$
expansion. However, the identification may be lost in the exponential
terms $\propto \exp{(-2E_q/\mu_{\rm hadr})}$.

%%%%%%%%%%%%%%%%
\subsection{Small Velocity expansion in $b\to c$ transitions}
%%%%%%%%%%%%%%%%%

Many simplifications arise for $b \to c$ semileptonic decays when
the velocity of the final state hadronic system is small with both initial
and final state quarks sufficiently heavy. The physics of
these SV transitions is also more transparent \cite{sv}. 
The SV case is a particularly relevant 
testground for duality violation: the OPE for total inclusive widths
relies on the expansion in the energy release $m_b\!-\!m_c$ rather than in
$m_b$ itself. In addition, the peculiarities of the four-fermion decay
interaction which manifest themselves in the high power $n\!=\!5$ of
the dependence on the quark masses, suggest \cite{five} that in
general the actual
hardness of the total widths can be further decreased down to only a
fraction of $m_b$.  
This {\it a priori} can boost effects of local duality violation, but
by the same token enforces the onset of the SV kinematics. Here we 
briefly describe the SV
regime for the total semileptonic width.

The heavy quark expansion of the semileptonic $b\to c$ width is
usually written as an expansion in $1/m_b$:
\beq
\frac{\Gamma_{\rm sl}(B\to X_c)}{\Gamma_0(b\to c)} =
c_0\left(m_c/m_b\right) + 
c_2\left(m_c/m_b\right)\, \frac{\mu_2^2}{m_b^2} + 
c_3\left(m_c/m_b\right)\, \frac{\mu_3^3}{m_b^3}+ \ldots
\label{80}
\eeq
where the coefficient functions $c_k$ are computed in the
short-distance expansion depending on $\alpha_s$, and on $m_c/m_b$ as
a parameter. There is 
no contribution of order $1/m_b$ -- the principal result of the 
OPE \cite{buv}. 
In the SV expansion we
rearrange terms in Eq.~(\ref{80}) using instead the parameters
$E_r=m_b\!-\!m_c$ and $m_b$; this implies that $E_r\ll m_b$ holds in 
the applications. The velocity of the final state hadrons does not
exceed $(m_b\!-\!m_c)/m_c \simeq E_r/m_b \ll 1$, and 
we have 
\beq
\frac{\Gamma_{\rm sl}(B\!\to \!X_c)}{\Gamma_0(b\!\to\! c)} \!=\!
N \cdot  
\left[1\!+ v \,\frac{\nu_2^{(1)}}{E_r^2} \!+ 
v^2\left(\frac{\nu_2^{(2)}}{E_r^2} \!+\! \frac{\nu_3^{(2)}}{E_r^3}\!+ \ldots
\right)\!
+ 
v^3\left(\frac{\nu_2^{(3)}}{E_r^2} \!+\! \frac{\nu_3^{(3)}}{E_r^3}\!+ \ldots
\right) \!+\ldots \right] ,
\label{82}
\eeq
$$
v\equiv \frac{E_r}{m_b} = \frac{m_b\!-\!m_c}{m_b}\;.
$$
Here $\nu_k^{(l)}$ are hadronic parameters of dimension
$\Lam^k$. The OPE relates them to the expectation values of local
heavy quark operators in the $B$ meson state. This expansion is the
nonrelativistic version of the OPE of Refs.~\cite{buv,bs}
kinematically similar to the neutron $\beta$-decay, improved by
accounting for the relativistic effects order by
order in velocity. 
The SV semileptonic width was considered to order $v^2$ by
Le~Yaouanc {\it et al.} \cite{orsaynew} in a toy model where the OPE
generalities become explicit.
 
Expansion (\ref{82}) exhibits a few peculiar features. First,
there are no nonperturbative corrections in the velocity independent
term --   
i.e., even at $m_b\!-\!m_c \to 0$, for large enough heavy quark
mass. The impact of strong
dynamics here is given solely by the short-distance factor
$N=1+\frac{3}{4}g_A^2$ with $g_A$ denoting the perturbative
renormalization of the (diagonal) axial current. 

Secondly, there is a unique term linear in $v$; it is related
to the quark spin and is absent for spinless quarks (or for vector-like
weak interactions).

The OPE explicitly performed for $\Gamma_{\rm sl}$ through order
$1/m_b^3$ gives concrete expressions for the terms $\sim 1/E_{r}^2$ and
$\sim 1/E_{r}^3$, with arbitrary powers of $v$ {\it via} the kinetic,
chromomagnetic and Darwin expectation values in $B$ meson.
The SV OPE also exhibits distinct features. Namely, the OPE series to a given
order in $v$  have only a finite number 
of terms\footnote{This follows
from the second representation in Eq.~(\ref{72}). In the SV expansion
the structure functions are polynomials in $\vec{k}$; integrating them
over $\vec{k}$ yields a finite polynomial in $k_0$ whose
nonperturbative expansion therefore has only a limited number of terms.} 
in $1/E_r$, although
this number grows with the power of $v$. In particular,
there is a single term to order $v^1$ as anticipated in
Eq.~(\ref{82}), which is given by $\mu_G^2$:
\beq
\nu_2^{(1)}= -\frac{10 g_A^2}{1\!+\!3g_A^2} \, \mu_G^2(m_b^2)
\label{84}
\eeq
(there are already seven terms through order $v^2$).

Returning to violations of local duality in the semileptonic decay
widths, it is likewise advantageous to analyze it as the difference
between the exact width and its OPE expansion, both considered to a
given order in $v$. 
This has an added convenience 
since then one does not have to address 
the general problem of 
defining an exact sum of the asymptotic expansion
with factorially growing coefficients $\nu_k$. In this respect, for any
particular term $\sim v^l$ the situation is similar to the case of
the $\tau$ decay width considered earlier. Similarly, the fact is evident
that duality violation is to be present at some level regardless of
details of dynamics.\footnote{The asymptotic nature of the expansion in
velocity Eq.~(\ref{82}) itself is an additional source of duality
violation. Since it is not instructive, 
we do not discuss it here.}

As mentioned above, duality is exact to order $v^0$. It is less
trivial {\it a priori} that the same holds even to the next order in
$v$: There is no difference between $\Gamma_{\rm sl}$ and 
$\Gamma_{\rm OPE}$ up to terms $v^2$ at {\it any} energy release! 
This can
be seen by noting that only the ``quasielastic'' widths $B\to D$ and $B\to
D^*$ must be considered. The chromomagnetic interaction
shifts masses in the same direction for $B$ and $D$;
therefore, this effect shows up only to ${\cal O}(v^2)$. However,
$\delta_{\mu_G^2} M_{D^*}=-\frac{1}{3}
\frac{m_b}{m_c}\delta_{\mu_G^2} M_{B}$, and this splitting is the only
source of ${\cal O}(v)$ corrections. (Recall that the phase space
is controlled by the fifth power of $E_{r}=vm_b$.) Exactly the same
contribution evidently emerges in the OPE to order $v^1$. 

The actual violation of local duality in $\Gamma_{\rm sl}$ emerges only
to order $v^2$ and is suppressed by at least two powers of
$E_{r}$. Since $\frac{v^2}{E_{r}^2}=\frac{1}{m_b^2}$ these effects
start with terms $\propto 1/m_b^2$; higher order terms can be
further suppressed already by factors $\mu_{\rm hadr}/(m_b\!-\!m_c)$. 
In actual $B$ decays $E_{r}\simeq 3.5 \GeV$, i.e.\ much higher than,
say, the $\tau$ mass; this underlies the strong numerical suppression of
the local duality violation in $\Gamma_{\rm sl}(B)$.

To order $v^2$ the total decay rate is given by a few moments of
the nonrelativistic SV structure functions. (The exact combination of
moments is determined by the nonrelativistic form of
Eqs.~(\ref{66}).) They are
computed, however, over a limited energy range. If the energy
release exceeds the mass of the heaviest charm state (the ``$P$ wave''
excitations contribute here), these moments exactly coincide with
their OPE expressions. In practice, there are always high enough
charm states, for instance, dual to perturbative gluon excitations of
charm. Therefore, the onset of local duality in the SV width is
directly related to the scale where nonperturbative contributions to
the sum rules are saturated and to onset of the perturbative regime
there, the dashed-dotted line in Fig.~3. We will refer to this figure
more than once later, 
and its meaning will be explained in more detail there.

%%%%%%%%%%%%%%%%%%
\section{Theoretical implementations}
%%%%%%%%%%%%%%%

As illustrated above, while the concept of quark-hadron duality 
can be formulated accurately whenever the OPE can be applied, it 
represents a complex phenomenon. 
Studying it in a concrete model would be of great help. However 
this represents a nontrivial task; for  
some aspects of the OPE and 
thus also of duality 
depend on subtle features rooted in the gauge nature of QCD.
Since 
those are typically not respected in naive quark models, 
doubts arise in the relevance of their numerical estimates of duality 
violations. Below we describe two other dynamic implementations of 
nonperturbative physics. While both are highly nontrivial, neither 
contains the full complexity of real QCD. Yet they are largely 
complementary in accounting for different facets of nonperturbative QCD. 
More details can be found in the already mentioned review by Shifman
\cite{shifio} or in the original publications referred to there.

%%%%%%%%%%%%%%%%
\subsection{Instanton model}
%%%%%%%%%%%%%%

Instantons provide a nontrivial dynamic realization of
nonperturbative physics generating full OPE series which, in principle,
can be evaluated to sufficiently high order. They are believed to be
relevant for the properties of low and intermediate energy physics
shaping a host of nonperturbative parameters of QCD in Euclidean
space. Models of this type assume that quarks and gluons propagate
not in the perturbative vacuum, but in the background of instanton
configurations of typical size $\rho_0$. The instanton density 
decreases fast for instantons of a smaller size. 
It is usually
assumed that the density of instantons is small enough compared to
$1/\rho_0^4$ in order to preserve a meaningful notion of 
individual instantons or to facilitate the computations. In this
approximation the instanton effects are proportional to their density
and strongly depend on their mean radius $\rho_0$.

The instanton ansatz is instructive in a number of aspects
related to duality and its violations \cite{inst}.  

(i) It has been explicitly shown to lead to duality violations 
in total semileptonic widths contrary to the often stated 
lore that while nonleptonic widths suffer from 
duality violations, semileptonic ones do not. The instanton 
calculus has demonstrated a quantitative rather than 
qualitative distinction in this respect between the two 
processes. 

(ii) The instanton ansatz illustrated that finite-distance 
singularities in the Green functions -- leading to divergences of the 
OPE and to violations of local duality -- are a common, rather than
exceptional feature of strongly interacting systems. Moreover, it 
showed that duality violations are not intrinsically tied to
confinement, contrary to what a historical perspective might suggest. 
At the same time, the instanton ansatz exhibits the
general features discussed above -- oscillations and fast decrease with
energy, strong suppression upon averaging, larger effects in
nonleptonic widths compared to the semileptonic ones, etc. 

(iii) 
On the practical side, it was shown that conventional instantons
cannot induce any appreciable duality violating effects in total
semileptonic widths of $B$ particles, regardless of uncertainties in
the model parameters. Even boosting up the possible instanton density 
leaves the effects below the permill level. If some appreciable
oscillation observed in $e^+e^-$ or $\tau$ decay distributions are 
rooted in 
such effects, the responsible nonperturbative configurations must have
significantly {\it smaller size.} While having a minor effect on 
Euclidean short-distance observables at energy scales around $m_b$,
they would be manifest at a lower scale $\sim m_c$ even in the Euclidean
domain.

%%%%%%%%%%
\subsection{'t~Hooft model}
%%%%%%%%%%%

The 't~Hooft model \cite{thooft}  -- a (1+1)-dimensional analogue of QCD with
$N_c\to\infty$ -- is an attractive theorist's 
laboratory in exploring various
complicated aspects of nonperturbative dynamics in QCD. Being 
solvable, it allows in principle deriving precise numerical values of
the model's counterparts of actual hadronic characteristics. Then it
is possible to confront them with the results of particular
approximations employed in real QCD and in this way to test their
viability.

This model is particularly appealing for studying duality and its 
limitations. It automatically respects the basic underlying features
of QCD related to gauge invariance, including its rigorous sum rules; 
{\it ad hoc} models typically fail in that respect. At
the same time, there is little `wiggle room' for adjusting various
parameters, so that the results are subject to smaller
interpretational bias. 

Since the final states in the 
't Hooft model consist of an infinite series of discrete narrow
resonances, one  expects this model actually to {\em maximize} 
local
duality violations. Additional considerations can be found in the
dedicated papers \cite{d2,d22}.

Since it is solvable, one can determine inclusive 
widths by calculating the rates for all available exclusive channels and sum
over them. 
Comparing the sum with the OPE result provides a direct test of duality.

This program has been performed for both
nonleptonic and semileptonic decays in the above papers, and full 
consistency with the OPE has been found for both semileptonic and
nonleptonic decays. In particular, the $1/m_Q$ terms forbidden by the OPE
are absent from the total widths.\footnote{It had been claimed in the 
literature (and most recently reiterated by Grinstein) that the OPE is
inapplicable to inclusive decay widths, based on the alleged presence 
of $1/m_Q$ terms in the total widths within the 't Hooft model as
deduced from some numerical computations. Those findings
were erroneous.}

At the same time, violation of local duality allowed by the OPE is
explicitly observed in the 't~Hooft model. The difference between the
actual width and its OPE expansion is oscillating and fades out fast
with the increase in energy. The decrease happens even {\it without}
including finite widths for the resonances or smearing in energy.
The degree of suppression, however, is non-universal and depends on the type
of the transition. 

The numerical aspects of duality in the semileptonic decays in the
domain of intermediate masses were analyzed in Ref.~\cite{lebur}. It
has been established that the prominent resonance dominance does 
not spoil
duality, nor delays its onset. The key fact is the saturation of a
few lowest sum rules: above this scale the amount of duality
violation is negligible. Violation of local duality in the
total semileptonic $B$ decay widths was found to be safely below the 
permill level even considering the possible uncertainties in
translating quark masses. It should be noted that in general the
nonperturbative effects in the 't~Hooft model are quite significant, in
particular, in exclusive heavy quark amplitudes. Inclusive decay
widths appear to belong to the class of observables most robust
against the effects of nonperturbative dynamics.

%%%%%%%%%%%%%%%%
\section{Application to $\Gamma_{\rm sl}(B)$}
%%%%%%%%%%%%%%%%%

A central phenomenological application of the heavy quark expansion 
in beauty decays is extracting the value of the CKM parameter 
$V_{cb}$. 
The total semileptonic width of $B$ mesons is a well measured 
quantity that can be treated theoretically through the OPE. 
Thus it is important to scrutinize the possible size of local
duality violation in  $\Gamma_{\rm sl}(B)$.

%%%%%%%%%%%%
\subsection{General features}
%%%%%%%%%%

While the exact mechanism driving
violation of local duality in QCD is not reliably known, 
it has many features which significantly limit its magnitude in 
semileptonic decays. We illustrate this 
first semiquantitatively using the 
general arguments of 
Ref.~\cite{varenna}, Sect.~2.5.3; in the next subsection we present 
a more specific estimate.

The underlying expansion parameter for $\Gamma_{\rm sl}(B)$ is the 
energy release $E_r=m_b\!-\!m_c\simeq 3.5\GeV$ rather than the safely large 
$m_b$.
Therefore, we can analyze the amount of
duality violation varying  $m_c$ from $0$ to $m_b$ while keeping 
$M_B$ fixed to its actual value:
\beq
\chi(E_r)= \frac{\Gamma_{\rm sl}(B\to X_c)}{\Gamma_{\rm OPE}(b\to c\,
\ell\nu)}-1
\;.
\label{90}
\eeq

For $m_c \!\to\! 0$, i.e.  
effectively for $b\!\to\! u$,  
the magnitude of duality violation is expected to be quite
negligible, since the energy release is plentiful. We shall 
conservatively assume it to be below $10^{-3}$,
even though all realistic estimates yield much smaller values
\cite{inst,lebur}. The opposite side of the interval -- $m_c \to m_b$ --  
deserves careful attention, presumably yielding the largest values of
$\chi$. 
One still has to maintain 
$m_b \!-\! m_c \gsim \mu_{\rm hadr}$ to keep the literal OPE series 
meaningful.\footnote{
It is curious to note
that even at $\Lam \gsim m_b-m_c \gg \frac{\Lam^2}{m_b} \sim 40\MeV$ 
duality holds up to terms $\sim \frac{\mu_{\rm hadr}^2}{m_b^2}$; 
assuming
$\mu_{\rm hadr}^2 \sim 0.5\GeV^2$ one would estimate this effect as only
a few percent.}

For $m_b\!-\!m_c \sim 1$ GeV one has $v\approx 0.2$ for the velocity of the 
$c$ quark and the SV expansion Eq.~(\ref{82}) becomes a valuable
tool. 
As explained in Sect.~4.3 duality is exact to orders ${\cal O}(v^0)$ 
and ${\cal O}(v^1)$. It is worth
noting that the leading nonperturbative term in Eq.~(\ref{82}) amounts
to about $15\%$ at $m_b\!-\!m_c=1\GeV$ ($5\%$ at the actual
$m_b\!-\!m_c=3.5\GeV$), and duality is still {\it exact} at this level!

Violation of local duality as measured by $\chi$ 
emerges first at ${\cal O}(v^2)$. 
The typical scale of the relevant nonperturbative effects can
be estimated by taking just the leading OPE term. With
\beq
\nu_2^{(2)}\simeq \frac{7}{4}\mu_G^2 - \frac{1}{2}\mu_\pi^2
\label{92}
\eeq
it is around $2\%$. In reality, the {\it duality violating}
component must be even further suppressed. 

Through order $v^2$ both the OPE terms and the individual decay
probabilities are determined by the SV amplitudes -- the transition
formfactors called $\tau_{\frac{3}{2}}$ and $\tau_{\frac{1}{2}}$ to
the excited states in the heavy quark limit (the exact definitions
can be found, for example, in the review \cite{ioffe}). They
have been well studied theoretically. The OPE parameters $\nu_2^{(2)}$,
$\nu_3^{(2)}$,... are given by the corresponding moments of the SV
structure functions, which are sums of terms like
$|\tau_k|^2\epsilon_k^l$. The very same sums appear in the total width
computed as a sum of the quasielastic ($B\to D, D^*$) and $P$-wave
decay probabilities. The only difference is that the actual decay
width includes in the sum only the kinematically allowed states, those
for which $\epsilon_k \lsim m_b\!-\!m_c$. Thus, if the SV sum rules were
completely saturated at a given energy release, duality violation
would have been totally absent at this level, and 
$\chi \lsim {\cal O}(v^3)$. 

The first $P$-wave states have 
excitation energy $\epsilon_1 \simeq
400\:\mbox{to}\:500\MeV$. In the 't~Hooft model, they almost completely
saturate the sum rules which implies minute duality
violation even at the minimal energy release. This probably does not 
hold for 
real QCD, where a significant contribution can be attributed to the states
with $\epsilon$ up to $700\,\mbox{to}\,800\MeV$, in particular for 
the higher 
moments. Let us note that in traditional analyses (say, in the QCD sum
rules) the sum rules are assumed to be saturated within ten percent
by the excitations below $1\GeV$. This would lead
to a similar minute duality violation at the sub-$10^{-3}$ level
already at $m_b\!-\!m_c \simeq 1 \GeV$.

We do not go that far; instead we 
assume 
$\nu_2^{(2)}$ to be merely $75\%$ saturated, and allow an even more
modest saturation of $50\%$ for higher moments. Then we arrive at 
\beq
|\chi (1\GeV)| \lsim 1\%
\label{94}
\eeq
A more refined bound can be obtained applying reasoning similar to
the one we employ below, Sect.~6.2, making use of general
properties of duality violation: one can relate the maximal magnitude of
the duality violation to a fraction of the branching fraction of 
the highest open threshold. Namely, to order $v^2$ we have 
\beq
\chi(E_r) = \frac{1}{m_b^2} Q^{(2)}(E_r),
\label{96}
\eeq
where -- analogous to $\tau$ decays -- $Q^{(2)}$ is an oscillating
function with a known threshold behavior at each $P$ wave state, and its
first few moments vanish. Although this results in a more stringent
bound, we do not present it here.

The terms of higher order in the velocity 
$v$ are suppressed from the very
beginning, $v^3 \simeq 0.01$ at $m_b\!-\!m_c=1\GeV$. Therefore, we
conclude that the magnitude of duality violation is at most a fraction
of a percent already at $m_b\!-\!m_c=1\GeV$.

Since the amplitude of duality violating oscillations decreases fast
with energy release, we see why it always emerges extremely small for
$b\to u$ transitions. With the amplitudes decreasing as a power of
energy release, the scale of duality violation at physical masses,
$E_r \simeq 3.5 \GeV$ is found well below a phenomenologically relevant
level. Some additional qualitative arguments including a comparison
with $\tau$ decay width can be found in Ref.~\cite{varenna}.

%%%%%%%%%%%%%%%%%
\subsubsection{Nonperturbative hadronic scale in semileptonic $B$ widths
and heavy quark symmetry in charm}
%%%%%%%%%%%%%%%

Fig.~3 will help to make our reasoning more transparent: it shows 
salient features of the typical hadronic mass distribution in 
$B \to X_c \,\ell\nu$.  

The canonical assumption is that the energy
scales above $1\GeV$ beyond $M_{D^*}$ belong to the 
perturbative regime. Yet based on the
QCD Lagrangian alone we could not rule out {\em a priori} that, 
for example, a prominent charm resonance existed with a mass 
above $4\GeV$ that is not shadowed by a depletion in the 
continuum hadronic mass spectrum nearby, and so altogether 
the total yield were not dual to the perturbative spectrum. If such an 
unorthodox resonance exceeded $M_B$ -- or were at least 
close to it -- this could lead to a poor convergence of the 
OPE and/or sizeable duality violation in $\Gamma_{\rm sl}(B)$. 
In the language of 
Fig.~3 it would mean that the resonance region $R$ extended up to 
or even beyond $M_B$ with little or no room for the perturbative 
domain $P$. 

Based on our whole experience with QCD this is not a likely 
scenario, but its consequences can be analyzed. Such 
a phenomenon would leave heavy footprints in other 
transitions involving charm as well. In particular it would drastically 
impair the calculability of the 
$B \to D^*$ formfactor even at zero recoil. Oddly 
enough, this connection is mostly missed in the literature.

The point is that even the Euclidean heavy quark expansions depend on
the same nonperturbative operators, and their expectation values are
correlated with the spectrum and the residues of the heavy flavor
resonances. Suppose we have a charm state $X$ with $M_X\simeq 4.5\GeV$
and the amplitude $\matel{X}{\bar{c}\gamma_\mu(1\!-\!\gamma_5)b}{B}$
such that its branching fraction constitutes, say, only a minute
$0.5\%$ part of $\Gamma_{\rm sl}(B)$. The velocity of such a hadron
being 
very small, it must be either an $S$-wave or $P$-wave
transition. In the heavy quark limit only $P$-wave amplitudes
vanishing at threshold survive,
\beq
\frac{1}{2M_B}\matel{X(\vec{v}\,)}{\bar{c}\gamma_\mu(1\!-\!\gamma_5)b(0)}{B}
\propto \tau_X \vec{v}\;.
\label{180}
\eeq
However, there are $1/m_c$ corrections which yield the amplitude
nonzero even at vanishing recoil, and they are of order unity at such
$M_X$. Using the nonrelativistic expansion of the zero-recoil
$\bar{c}b$ current one finds (see \cite{optical}, Sect.~VI) that such
amplitudes are generally given in terms of $\frac{M_X\!-\!m_c}{m_c}\,
\tau$, up to spin-related factors. Therefore, we can simply assume the
amplitude a constant at small $\vec{v}$, and approximately equate it
to the corresponding SV amplitude $\tau$ in the heavy quark limit. 
The partial decay width can then be estimated as 
\beq
\Gamma_{\rm sl}(B\to X) \approx \frac{G_F^2 |V_{cb}|^2}{15\pi^3}\,
|\tau|^2 (M_B\!-\!M_X)^5\;,
\label{184}
\eeq
that is, say, even for $M_B\!-\!M_X \simeq 1\GeV$ we would need
\beq
\tau_X^2 \simeq 0.2
\;.
\label{186}
\eeq

However, let us look now at the contribution of such nonperturbative states
to the sum rules, see, e.g. \cite{ioffe}. The increase in the IW
slope through the Bjorken sum rule is given by
\beq
\delta_X\varrho^2 \approx 3\tau_X^2 \approx 0.6
\label{188}
\eeq
and is, in principle, tolerable keeping in mind the tentative nature of
the estimates. However, everything is different already for $\La=M_B\!-\!m_b$
and, in particular, for $\mu_\pi^2$:
\beq
\begin{array}{lll}
\delta_X\La \!\!&\approx 6\tau_X^2 \epsilon_X \!\!& \approx \;\,3\GeV\\
\delta_X \mu_\pi^2 \!\!&\approx 9\tau_X^2 \epsilon_X^2 \!\!& \approx 11\GeV^2\;,
\end{array}
\label{190}
\eeq
not to mention the Darwin term and other higher-order operators. On
the other hand, the literal value of $F(0)$ in the $1/m$ expansion
shifts at least by $-1\%$ for an increase in $\mu_\pi^2$ by $0.12
\GeV^2$ (this is a {\tt model-independent} bound \cite{vcb}). It is
then obvious that through the {\it exact} dispersion relations and the OPE the
expansion of the $B\to D^*$ amplitude is orders of magnitude more
sensitive to the high-momentum nonperturbative dynamics than the potential
duality violation in the total semileptonic width. 

The above estimates can be cast into a more accurate form using the
whole set of the heavy quark sum rules including new exact sum
rules \cite{newsr}. We have partially incorporated the latter in
Eqs.~(\ref{188}), (\ref{190}) and (\ref{184}) assigning approximately
equal values to the corresponding $\tau_{\frac{1}{2}}$ and
$\tau_{\frac{3}{2}}$ for the highly excited states. If this were
violated, one would completely destroy the exact relation for the $B$
meson spin and got an unacceptedly large value for the
chromomagnetic operator $\mu_G^2$. However, in view of the obvious
trend of numbers, such an improvement seems superfluous. 

The key point illustrated by the above consideration is
transparent. Physics of power corrections in individual $b\to c$
semileptonic transitions is strongly correlated with the saturation of
the heavy quark sum rules, since the former include the expectation
values given, for example, by the moments of the SV structure
functions. Allowing appreciable genuine nonperturbative effects in
the heavy quark hadronic system with excitation energies $\mu$
exceeding a couple $\mbox{GeV}$ would dramatically upset the $1/m_c$
expansion leading to the higher order terms scaling as powers of
$\mu/m_c$ instead of the naive $\Lam/m_c$, while only moderately
affecting the total semileptonic $B$ width. This simple interrelation
was put forward already in the review \cite{rev}, but seems to be
missed up to now by most of the heavy quark theorists speculating
about  duality violation in the total semileptonic widths.

%%%%%%%%%%%%
\subsection{A realistic estimate of duality violations in 
$\Gamma_{\rm sl}(B)$} 
%%%%%%%%%%%%%%

The above considerations were intended to illustrate the scale of 
rather model-independent factors suppressing violations of local 
duality in inclusive semileptonic widths. Augmenting them with 
explicit estimates of the power terms and the asymptotics of
the nonperturbative formfactors yields an actually much 
stronger suppression. The main lesson here parallels what has been
derived in the 't~Hooft model: the usual resonance-related duality
violation becomes negligible as soon as the characteristic scale of their mass
gap is passed \cite{lebur}. This scale 
$\mu_{\rm hadr}$ is believed to be about $0.7\,\mbox{to}\, 1\GeV$ in QCD. 

Let us take a look at Fig.~3: $\rho(M_{X_c})$ corresponds to the structure
functions $w$ of the $b$ quark inside the $B$ meson 
appropriately integrated over the spacelike momentum. It is related to the 
total decay width as a function of the energy release with the lepton
phase space factored out. For clarity, we actually multiplied
this quantity by the factor $M_{X_c}\!-\! m_c$ to make it
approximately constant in the perturbative regime. 
The domain $R$ 
below $M_{D^*}+\mu_{\rm hadr}$ (i.e., $E_r\lsim \mu_{\rm hadr}$) 
is usually referred to as the resonance region, while the one above it
is viewed as described by perturbative dynamics. 

\thispagestyle{plain}
\begin{figure}[hhh]
\vspace*{1mm}  
 \begin{center}
 \mbox{\epsfig{file=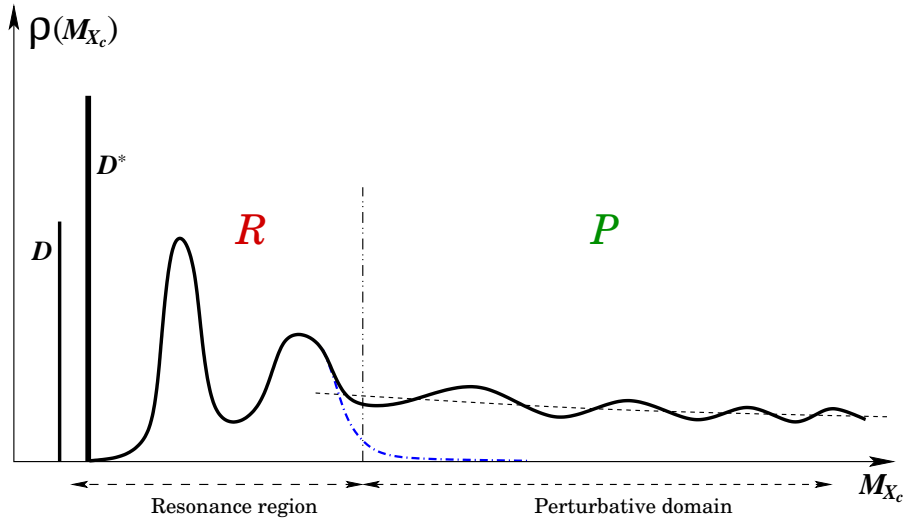,width=12cm}}
 \end{center}\vspace*{-6mm} 
\caption{ \footnotesize
Plot illustrating different strong interaction
domains. $\rho(M_{X_c})$ stands for a generic $b\to c$ structure function
of a beauty hadron, related to the observable distribution over
hadronic final state mass up to kinematic factors. 
{\normalsize {\color{red}$R$}} 
\,marks the `standard'
nonperturbative domain dominated by low-lying resonances. In the quark
models the distribution fades out very fast above it in the 
{\normalsize {\color{green}$P$}} domain
({\color{blue}dash-dotted} tail), but is
larger in actual QCD where it is dual to the parton processes with hard
gluons. The decay distribution there may also oscillate around the
perturbative prediction shown by the dash line, up to higher energy
scale, with a larger interval of local duality. The excitation energy 
$M_{X_c}\!\!-\!M_{D^{(*)}}$ at
the borderline between the two domains is generically denoted
$\mu_{\rm hadr}$. 
}
\end{figure}

Our preceding discussion concentrated on the duality violation which
is associated with the resonance domain $R$. In a sense, we implied
that the perturbative contributions were absent, and the distribution in
Fig.~3 nearly vanishes above $M_{D^*}+\mu_{\rm hadr}$;  this is 
similar to the
situation in the 't~Hooft model where the production of highly
excited states with masses above $m_c+\Lam$ is strongly power 
suppressed,
and the principal nonperturbative effects originate from and can be
related to a few lowest resonance states in the domain $R$.  If the
resonances were totally absent above the dash-dotted borderline 
$M_{D^*}+\mu_{\rm hadr}$,
the actual hadronic width would exactly coincide with its OPE
expansion for $E_r >\mu_{\rm hadr}$.
The numerical studies of the duality in the 't~Hooft model
\cite{lebur} were undertaken just to elucidate how the prominent resonance
structure in the nonperturbative domain would affect the convergence
of the 	`practical' OPE as we approach this borderline and if it could
delay the onset of accurate local duality.  

In real QCD, however, the relevant excitation probabilities decrease
not powerlike as in simple models, but much slower due to the
perturbative effects associated with the emission or exchange of hard
gluons, as shown explicitly in Fig.~3. Physics of such processes has
not been included so
far. In actual $B$ decays we are safely in the perturbative domain, as
in Fig.~3, if we place $M_B$ close the right edge of the plot. As  
suggested in Refs.~\cite{d22,lebur}, an appreciable
violation of local duality could possibly emerge from this type of hadronic
excitations. We present its estimate in this section.

The instanton effects addressed in Ref.~\cite{inst} would likewise
populate the domain above the resonance region. However, they lead to
oscillations which decrease very fast in magnitude and can be
neglected. They would not be visible in the relevant part of the
distribution in Fig.~3.

The perturbative corrections {\it per se} do not vitiate the OPE even
though they generally lead to divergence of the spectral
functions moments. The short-distance expansion, instead of being regular
order by order at large energies, acquires logarithmic factors or, more
generally, non-integer powers of energy yielding continuous imaginary
parts. It does not generate any duality violation by itself as is most
obviously illustrated by finite-order perturbation theory: with
the parton processes yielding smooth continuous spectra of states,
there is always exact duality at any energy, regardless of the
presence of kinematically allowed and forbidden states. Since all the
decay probabilities are smooth here, they coincide with their smeared
averages, and there is no room for violation of local duality. The
spectral density $\rho(M_{X_c})$ as it comes out perturbatively is
shown in Fig.~3 by the dashed line. 

However, the excitations of gluonic degrees of freedom may, in
principle, exhibit a certain resonance structure. If the
typical energy scale for unfreezing  gluonic degrees of freedom is
well below the energy release, our general discussion in the previous
sections applies, and duality violation can be neglected. On the other
hand, the mass gap for the transverse gluon degrees of freedom can be
larger than in the $q\bar{q}$ channels. Then, as conjectured in
Ref.~\cite{d22}, one would observe {\em two} scales in the
onset of local duality: it first sets in rather early,
however only approximately with a typical accuracy $\sim \alpha_s/\pi
\sim 10\%$; one needs to ascend to higher energies characterized by the
``gluonic'' mass gap to achieve full duality. 

In this scenario the resonance structure can be observed at a
suppressed rate up to higher energies, and its average is dual to the
probabilities of the gluon bremsstrahlung, see Fig.~3. In  
semileptonic heavy
quark decays the latter rates decrease slowly with energy compared to
a powerlike suppression of exciting the valence quark-antiquark
mesons. Therefore, this would dominate the local duality violation at
intermediate energies. Here we present an estimate of the possible
magnitude of these effects; the resulting upper bound is to a large
extent model-independent. To maximize the possible effect, we replace
the continuous oscillating behavior by a comb of $\delta$-functions
representing infinitely narrow resonance; however, upon average they
reproduce the perturbative distribution. 

The basic idea for this estimate goes back to Ref.~\cite{d2} where it was
applied to the $\tau$ decay width. It makes use of the following
facts:\\
$\bullet$ The average of $\Gamma_B-\Gamma_{\rm OPE}$ as a function
of mass vanishes.\\
$\bullet$ $\Gamma_{\rm OPE}$ is a smooth function given by a 
series in $1/m_Q$, while $\Gamma_B$ has known threshold
singularities. Each partial decay width is a smooth function above 
threshold.\\
$\bullet$ The principal thresholds are assumed equally spaced at
large masses, and the corresponding transition amplitudes follow a
power-like scaling. Then 
$\frac{\Gamma_B-\Gamma_{\rm OPE}}{\Gamma_{\rm OPE}}$ is, 
to leading order in $1/m_Q$, a periodic function up to an overall
suppression factor $1/m_Q^k$ with some index $k>0$. \vspace*{.5mm}\\
This allows to determine the actual leading order behavior of the
duality violation. The index $k$ is given by the (maximal)  
rate of the last open channel relative to the total decay rate -- it
also scales like $1/m_Q^k$. 

It should be noted that the exact expression for $\Gamma_{\rm OPE}$
generally requires additional definition since the series are
asymptotic. The best standard treatment (in mathematical applications)
employs truncating them at the optimal order growing with energy (or
even using a Borel-type resummation if the analytic properties are
known {\it a priori}). Instead, we adopt a cruder, but
more physical definition of the truncated OPE series: we simply
discard the OPE terms with powers suppressed beyond $1/m_Q^k$. That is,
we keep only a fixed number of powers, the first few terms in the OPE.
Although including higher terms may further improve the
approximation, it is superfluous for our purposes and is not adequate
having in mind the reality of the practical applications. 

To estimate the expected amount of duality violation, we
need the threshold behavior of the resonances and the strength of the
threshold amplitudes. In principle, in the heavy quark limit the least
suppressed are the transitions to the $P$-wave states with 
amplitudes $\matel{P_n}{J_\mu}{B} \propto \vec{v}$. This yields
structure functions $w_1$, $w_2$ proportional to  
$\vec{k}^{\,2}\propto \Delta^2$ where $\Delta=m_b^2\!-\!M_n^2$
determines the  energy above the
threshold of the resonance.  
However, duality violation is governed
by the transitions to resonances with $M_X\!\simeq \!M_B$, 
for which the heavy
quark limit is not applicable. 
We should  therefore assume that the threshold amplitudes
do {\it not} vanish. The minimal possible threshold suppression
comes from the phase space $|\vec{k}\,|$ and from the lepton tensor $k^2
g_{\mu\nu}\!-\!k_\mu k_\nu$. Upon integrating over $k^2$ in Eq.~(\ref{66})
this gives the  leading threshold contribution\footnote{As has been
emphasized  before, duality
in the total widths depends on the amplitudes at $k^2\!=\!0$, i.e.\ on the
threshold behavior of the total (rather than differential)  width. 
Therefore, $|\vec{k}\,|\sim
\Delta$, and not $\sqrt{m_b\Delta}$.} $\propto
\Delta^{5}$. Thus, the relevant case is
$\gamma\!=\!5$ in the notations of the Appendix. (We also note that 
dimensional counting yields $k\!=\!6$ for
the scaling of the relative threshold width:
$\frac{\delta_n\Gamma}{\Gamma} \sim \frac{\Delta^5}{(m_b^2)^6}\,$.)

Using the analysis given in the Appendix, we find 
\beq
|\chi| \sim 
0.004 \frac{{\rm BR}_{\rm last}}{{\rm BR}_{\rm sl}(B)}\;,
\label{110}
\eeq
where ${\rm BR}_{\rm last}$ denotes the branching ratio of the last
fully open principal resonance in the perturbative continuum. The
latter rate is to be evaluated as yielding, upon averaging over the
successive resonance mass gap, the perturbative decay rate:
\beq
\frac{1}{6}{\rm BR}_{\rm last} \simeq 
\frac{\int_{M_B^2\!-\!\Delta_R}^{M_B^2} 
\frac{{\rm d}\Gamma^{\rm pert}} {{\rm d}M_X^2} {\rm d}M_X^2 }{\Gamma_B}
\;,
\label{112}
\eeq 
where $\Delta_R$ denotes the resonance spacing. The 
${\cal O}(\alpha_s)$
perturbative spectrum $\frac{{\rm d}\Gamma^{\rm pert}} {{\rm d}M_X^2}$
is explicitly known. Instead, we can use the simple dipole radiation
approximation where the gluon radiation probability is given by \cite{dipole}
\beq
\frac{1}{\Gamma}\frac{{\rm d}\Gamma^{\rm pert}} {{\rm d}M_X^2}  
\approx  \frac{8\alpha_s(\omega)}{9\pi} v^2  \frac{1}{2M_X (M_X-m_c)}
\;,
\label{114}
\eeq
with $\alpha_s^{(d)}$ the dipole radiation coupling known to two
loops and $v$ the effective velocity of the charm system (recall that $q^2
\simeq 0$):
\beq
v\lsim \frac{M_B^2-M_X^2}{2M_X^2}\;.
\label{116}
\eeq
Such a relation would be only approximate in the actual $B\to
X_c\,\ell\nu$ decays since $M_D$ is not large on the scale of the
radiated gluon energy. Therefore, we adopt an even simpler direct
phenomenological bound. Since the invariant mass near the threshold
$M_X \!\simeq 5 \!\GeV$ is significantly larger than $M_D$, we assume the
usual linear in mass square spacing of the successive
resonances. Next, we generally attribute the domain of $M_X$ above
$M_{\rm pert}\!\simeq\! 2.5\GeV$ to the perturbative excitation
probability. The integrated rate in the numerator of Eq.~(\ref{112})
is then estimated as the total perturbative fraction 
$\Gamma_{\rm sl}^{\rm pert}$ multiplied by the ratio of the mass square
gap to the total length of the perturbative interval in $M_X^2$:
\beq
\frac{1}{6}{\rm BR_{last}} = 
\frac{\Delta M^2}{M_B^2\!-\!M_{\rm pert}^2}
{\rm BR_{sl}} \frac{\Gamma{\rm _{sl}^{pert}}}{\Gamma_{\rm sl}(B)}
\label{118}
\eeq
This is clearly an overestimate, since the high-mass end of the
spectrum is suppressed by the decrease both in the lepton phase space and
in the effective QCD coupling at large energies. Yet this is
justified since we strive to obtain an upper bound for the duality
violation. 

For usual quark-antiquark states the gap $\Delta M^2$ constitutes
about $1\GeV^2$. As argued in the beginning of the section, 
appreciable duality violation can emerge if this mass gap is larger
for gluonic excitations. Therefore, we take $\Delta M^2$ 
as large\footnote{Clearly, such a high scale could not but affect
noticeably the perturbative corrections to $R_\tau$.} as $5\GeV^2$.

Assembling all pieces together, we arrive at 
\beq
|\chi| \lsim 0.006  
\frac{\Gamma{\rm_{sl}^{pert}}}{\Gamma_{\rm sl}(B)} 
\label{120}
\eeq
The fraction of the perturbative semileptonic width 
$\frac{\Gamma{\rm_{sl}^{pert}}}{\Gamma_{\rm sl}(B)} $ constitutes
about $10\%$. This direct experimental estimate is in  
agreement with the theoretical value derived from Eq.~(\ref{114}). 
Thus, taken at face value, we arrive at a {\it maximal} amplitude of
the duality violation in this mechanism only at a permill level:
$$
|\chi| \lsim 10^{-3}\;.
$$

To be conservative, we can increase this estimate by a factor of three
to five
making up for the approximations we have made, most notably using the
asymptotic expressions and neglecting the subleading in $1/m_b$
components in the amplitudes. We still end up with the amount safely
below a half percent. Alternatively, we can model the
preasymptotic effects by assuming softer threshold suppression. The
significant numerical suppression of $\chi$ comes from the
high fifth power of the threshold energy gap. This power properly 
determines the eventual suppression of the duality violation; however,
with high power one may have to pass more resonances before the asymptotic
counting rule sets in to work. To account for this, we can reduce the
effective threshold index $\gamma$ by a factor of $2$. Taking
$\gamma\!=\!5/2$ (this would, in a sense correspond to considering the
massive lepton pair contribution which may dominate in the domain of 
intermediate energies) we would have a softer numerical suppression
factor $|\tilde\chi_0|$ still yielding $|\chi| \lsim 1.5\cdot 10^{-3}$.

Since the resulting numbers may seem to emerge surprisingly small, it
is worth to address the qualitative breakdown of the suppression
factor. The most obvious factor is the overall perturbative suppression, 
$0.1\:\mbox{to}\:0.15$. Since the thresholds at $M_X \!\approx 
\!M_B\!\simeq
\!5\GeV$ lie in the perturbative domain at least qualitatively, such a
suppression is unavoidable. An additional factor is the fraction of this
width attributable to a single channel. It is rather uncertain,
however it emerged only $0.25$, and it clearly cannot
significantly exceed this amount.

The most significant suppression about $1/40$ comes 
from the nature of the duality
oscillations, the most general fact that duality violation can only
oscillate around the average predicted by the OPE. While the related
suppression is obvious as long as the theoretical applicability of the
OPE to the inclusive widths is not challenged, its magnitude may not be
fully appreciated {\it a priori}. In fact it is quite significant and
rests simply on unmodifiable mathematical constants. Let us
illustrate this fact. 

Suppose we start with the most singular meaningful threshold behavior
given by $(E\!-\!E_n)^0\,\theta(E\!-\!E_n)$. Simply subtracting the 
average already
makes the local duality violation suppressed by a factor of $2$ -- it is
described by the function $(\frac{1}{2}\!-\!x)$, $0\!\le \!x<1$. Increasing the
threshold exponent $\gamma$ by unity would amount, according to
Eq.~(\ref{108}) to roughly multiplying the maximal deviation by a
factor $(\gamma+1)/2\pi$, with $2\pi$ simply being the ``wave vector''
of the lowest Fourier mode on the unit interval. Thus, already for $\gamma=1$
one has the numerical suppression $1/4\pi \simeq 0.1$, and with 
$\gamma=2$ it becomes $1/4\pi^2 \simeq 1/40$.

It is evident that these principal suppression factors are
unavoidable as long as one has entered the higher resonance domain. With the
numerical suppression for the threshold behavior $\sim (E\!-\!E_n)^1$
already amounting to more than an order of magnitude, it is clear that
no effect exceeding a fraction of the percent can be identified with
such a mechanism. 

As explained above, this conclusion, although not having the status of
a theorem, nevertheless rests on quite general arguments and relies only on
very mild assumptions. 

A prominent effect which typically further suppresses the duality violating
oscillations is the finite resonance width, which has been completely
neglected. We note that due to suppressed threshold behavior
of the semileptonic widths, the oscillations are already rather
smooth. Therefore, the effect of the resonance mass smearing is far less
radical than, say, in $e^+ e^-$ annihilation (the case similar to
$\gamma\to -1$ in our notations) (cf.\ Ref.~\cite{bsz}), as long as
the resonances remain reasonably narrow, $\pi\Gamma_n \ll \Delta
M_n$ -- the damping of the oscillations is given by
$\exp{(\mbox{-}\frac{\pi\Gamma_n}{\Delta M_n})}$.

%%%%%%%%%%%%%
\section{Comments on the literature}
\label{LIT} 
%%%%%%%%%%

Calculability of the inclusive decay widths of heavy flavor hadrons in the
short-distance $1/m_Q$ expansion has been questioned from the time when the
OPE-based methods were elaborated and applied to quantify the
nonperturbative effects in $b$ hadrons \cite{buv,bs,WA}. The
arguments in the 1990's simply contended that the OPE as an expansion in
powers of $1/m_Q$ is not applicable to the total decay widths in
presence of confinement. However hardly any concrete evidence was
offered for such claims nor any faults in the derivation of the OPE
pointed out.  

After the validity of the OPE for decay widths had been illustrated 
explicitly in a number of nontrivial cases, the prevailing
terminology has gradually changed, but not necessarily the substance of the
criticism. It is raised by referring to ``quark-hadron duality'' as
an ``additional hypothesis''. Yet closer scrutiny reveals that
whenever concrete objections are formulated, the alleged effects 
are a violation of the OPE rather than of duality. The arguments thus
do not differ much from claims in Ref.~\cite{gl} of disproving the OPE for
inclusive widths based on numerical computations of the
decay widths in the 't~Hooft model. As is well known, a direct
analytic summation of the decay widths in the 't~Hooft model
demonstrated full agreement with the calculated OPE
\cite{d2,d22}. 
Here we briefly comment on more recent
publications referred to as ``supporting a possible sizeable source of
errors related to the assumption of quark-hadron duality''
\cite{barberio} and raising doubts in the ``quark-hadron duality {\it
ansatz}'' \cite{barberio}.

%%%%%%%%%%%%%%
\subsection{``Duality violation'' through `inapplicability' of the
OPE}
%%%%%%%%%%%%%%

The first paper \cite{isgur16} considered a toy model for the
semileptonic $b\to c$ width in the SV limit where the energy release
barely reaches the resonance domain. As mentioned above, the SV limit
is a convenient playground for inclusive widths. However,
to gauge the validity of duality 
the width obtained in the model was compared to the parton-level 
free quark width rather than to the OPE series -- i.e., taken without any 
calculable nonperturbative corrections. Even the 
constraints for $M_B\!-\!m_b$ following  from the ``optical'' sum rule
\cite{volopt} were not observed. (It is worth noting that in a 
few studied examples
including the leading OPE corrections decreased the magnitude of the 
difference by more than an order.)
Yet the difference that was found was interpreted as a violation of the
OPE in the cases with
`hard-confined' quarks.\footnote{One of us had discussions with the
author about these matters after Ref.~\cite{isgur16} appeared, and we
have reasons to believe that he is not fully behind such a claim
anymore.} It is difficult to accept such an interpretation 
  since the OPE {\it does predict} such a systematic difference! 

Another recent paper \cite{orsay17} also considered SV
semileptonic decays in a toy model. It is more elaborate 
and addresses higher powers in the expansion. It claimed to identify a
new mechanism of duality violation associated with would-be
contributions to the inclusive width from resonances
located above
the kinematically allowed mass range in the final state. 
One should keep two facts in mind: Firstly, this claim 
actually refers to an explicit {\it violation of
the  OPE} and, as such is not relevant to the violation of local
duality {\it per se} as it has been consistently defined. 
Secondly -- contrary to how the paper is sometimes quoted -- 
it does not claim any numerically significant effect. 

The basic idea behind the conclusion of Ref.~\cite{orsay17} is
actually the same that drove the thrust of Ref.~\cite{isgur16} -- that
the OPE wrongfully picks up contributions associated with 
transitions to states kinematically forbidden at a given $b$ quark
mass. It happens since the OPE allegedly is insensitive 
to the threshold factors
$\theta(M_B\!-\!M_n)$ expressing, in a sense, the conservation of
energy in the on-shell processes. 
For naively $\theta(M_B\!-\!M_n)$
can be dropped for any particular $n$ at $m_b\to\infty$, or at least
is not literally expandable in $1/m_b$ or $1/E_r$. 
In the SV approximation the partial
width of a state $P_n$ is proportional to $(M_B\!-\!M_n)^7$
(generally, to an odd power $2N+5$ of $(M_B\!-\!M_n)$ in the model of
Ref.~\cite{orsay17}). Therefore, the extra ``duality-violating''
contribution emerges always negative, and the actual width,
accordingly, would have always exceeded the OPE expression, by an
amount fading out as a power of energy release. 

The last simple observation shows that this would signify a
{\it violation of the OPE} for the width rather than a
manifestation of local duality violation which has to oscillate
around zero. The total absence of such an effect is actually 
easily illustrated in a complementary way. For one could apply the very
same reasoning to weakly coupled non-confined quarks (which is how
they appear in perturbation theory). 
The spectrum of ``hadronic
excitations'' is then continuous, the `kinematically forbidden' states
are there at arbitrary energy release, and they even are not
power-suppressed, only by powers of the coupling $\alpha_s$ (for more
details, see \cite{blmope,varenna,dipole,ioffe}). According to
Ref.~\cite{orsay17}, the OPE would completely fail --
while everyone would agree it works perfectly in this 
case. Actually, here the violation of
local duality would totally vanish since there are no thresholds and
the perturbative diagrams are smooth.

Since Ref.~\cite{orsay17} is clearly formulated, it is not difficult
to identify the subtlety which led to the erroneous conclusion. It
lies in the fact that there was no actual OPE performed, but instead a
simplified version was used which is only illustrative in elucidating
physics behind the OPE results for the decay widths and its relation
to the sum rules for the heavy quark transitions \cite{optical}. 

It can be easily shown that the $1/E_r$ expansion adopted in
Ref.~\cite{orsay17} coincides, term by term with the OPE if the masses
of all final state excitations do not exceed a certain mass
$(M_D +\mu_0)$. However, in this case Ref.~\cite{orsay17} would observe
exact vanishing of duality violation at $E_r \!\ge \!\mu_0$, in full
accord with our general arguments. We also note that the OPE in
$1/E_r$ for the transition amplitude would have had a finite radius of
convergence given by $1/\mu_0$; at $E_r \!<\! \mu_0$ the OPE series simply
diverges, even though the formal expansion of the width itself has
only a few terms. 

If the residues of the high-mass resonances do not vanish completely
but, nevertheless are exponentially suppressed,
\beq
\rho_n^2 \sim {\rm e}^{-c n^\beta}\;,
\label{140}
\eeq
the `practical OPE' series still coincide with the expansion of
Ref.~\cite{orsay17}. However, the ``condensates'' grow exponentially,
and the series for the transition amplitude is only asymptotic, much
like in the illustrative cases considered in
Refs.~\cite{shifdual,d2}. What is
crucial, however, is that the possible duality violation is
exponentially suppressed as well, like in any other Euclidean
quantity. There are no specific violations of {\it local duality}.

A definite violation of the OPE was found in Ref.~\cite{orsay17} in
the case where the residues $\rho_n^2$ fade out only powerlike,
\beq
\rho_n^2 \sim (M_n\!-\!M_D)^{-l}\;, \qquad M_n\!-\!M_D \sim n^{\alpha}\;.
\label{142}
\eeq
However, the employed expansion in this case does {\it not} coincide with
the OPE starting just the power $1/E_r^k$ with $k=l/\alpha+1$
where the discrepancy was found. Actually, the expansion simply cannot
be extended to this order since the corresponding moments of the SV
structure functions would diverge even in the toy model. The actual
OPE, however still exists at this level, and yields the right result!
Let us briefly explain this.

In the OPE, one considers the transition amplitude ${\cal A}(\omega)$
of Eq.~(\ref{52}) and expands it in $1/(\omega\!-\!E_r)$. For semileptonic
decays one can use the representation of Eq.~(\ref{74}) to get closer to
$T(q_0,\vec{q}\,)$ of Ref.~(\cite{orsay17}), and again  the
large-$\omega$ expansion is examined. This corresponds to the
asymptotic expansion of $T(q_0,\vec{q}\,)$ in $1/(E_r\!-\!q_0)$. It is an
easy exercise to check that to a first few orders in $1/(E_r\!-\!q_0)$ the
amplitude has the expansion 
\beq
T(q_0,\vec{q}\,) = \frac{1}{E_r\!-\!q_0} + \vec{v}^{\,2} \left(
\frac{c_2}{(E_r\!-\!q_0)^2} + \frac{c_3}{(E_r\!-\!q_0)^3} +
\frac{c_4}{(E_r\!-\!q_0)^4} + ...
\right)\;.
\label{144}
\eeq
Integrating the discontinuity of this expansion over $q_0$ would
exactly reproduce the moments through order $k$. (The support of
$\Im T$ for the expansion above lies at $E_r\!-\!q_0\!=\!0$, so 
one can integrate
over the actual physical interval, or formally till infinity.) 

However, the terms starting with $1/E_r^{k}$ become different: they
acquire nonanalytic pieces like 
\beq
\frac{\ln{(q_0\!-\!E_r)}}{(q_0\!-\!E_r)^{r}}\;, \;\;
\frac{1}{(q_0\!-\!E_r)^{r+p}}\;, \;\;
\frac{\ln^n{(q_0\!-\!E_r)}}{(q_0\!-\!E_r)^{r+p}}
\label{146}
\eeq
with some (generally, non-integer) powers of $E_r\!-\!q_0$ or
logarithms. Their discontinuity spans all the way to infinity and
simply cannot be integrated over the whole half-axis. In the OPE we integrate
it up to the proper threshold value $q_0\simeq m_b\!-\!m_c$, rather than
to infinity as in Ref.~\cite{orsay17}. Thus, in the actual OPE the
difference $\Gamma(B)-\Gamma_{\rm OPE}$ is {\it not given} by the
integral over the contour $C_L$ along the kinematically
forbidden range of masses, as was {\it ad hoc} {\tt postulated} {\it
ab initio} in the paper. 

Speaking concretely, the ansatz of Eq.~(\ref{142}) 
$$
\rho_n^2 \simeq r (M_n\!-\!M_D)^{-l}\;, \qquad M_n\!-\!M_D 
\simeq m\, n^{\beta}\;
$$
through dispersion relation leads to the following asymptotic term in
$T(q_0,\vec{q}\,)$:
\beq
\vec{v}^{\,2}\: \frac{r}{m^{\frac{1}{\beta}}}\:
\frac{\pi}{\beta\sin{(l\!-\!\frac{1}{\beta})\pi}}\:
 \frac{1}{(q_0\!-\!E_r)^{l-\frac{1}{\beta}+1}}
\label{147}
\eeq
(at physical $q_0$ in the decay $q_0\!-\!E_r <0$);
at integer $l\!-\!\frac{1}{\beta}$ it becomes 
$$
\vec{v}^{\,2}\: \frac{r}{\beta \,m^{\frac{1}{\beta}}} \;
\frac{\ln{(q_0\!-\!E_r)}}{(E_r\!-\!q_0)^{l\!-\!\frac{1}{\beta}+1}}\;.
$$ 
Moreover, the residues $\rho_n^2$ fade out powerlike only since a large
momentum scaling with $M_n\!-\!M_D$ can be exchanged there with the
light cloud. Correspondingly, the related terms are not missed and do 
appear in the short-distance OPE of the heavy quark Green function 
in the form of the high-dimension operators, with the coefficients
containing the gluon coupling. 

It is not difficult to check explicitly that the actual OPE does
correctly reproduce the terms $1/E_r^{k}$ (and a few more) in the
exact width $\Gamma(B)$. Instead of presenting the straightforward
derivation, we give here a simple heuristic argument which is more
than convincing. The point is that the amplitude given by the leading
term(s) in expansion (\ref{146}) is very similar to the effect from 
usual gluon perturbative corrections with nonconfined quarks and
gluons, however with the `coupling' suppressed by a corresponding
power of the energy. As explained above, nobody expects the OPE to
fail and local duality to be violated in such a ``non-resonant'' 
situation. This simply
means that going through the formal OPE machinery, viz.\ integrating
the imaginary part of the asymptotics of the transition amplitude is
bound to yield exactly the averaged contribution of the high hadronic
thresholds. The average difference between $\Gamma_{\rm OPE}$ and
$\Gamma(B)$ would vanish. \vspace*{2mm}

{\tt To summarize:}\, the observed violation of duality of
Ref.~\cite{orsay17} has its root simply in the fact that the actual OPE was
replaced by a procedure differing just in terms found
as an alleged violation of duality. 

We note that the most recent paper by the same authors 
\cite{orsaynew} appeared
while this Vademecum was in writing, dedicated to the question
of duality. It does not anymore claim establishing
duality violation at the considered level. We view this as an
indication that the authors do not stand behind the mechanism of
duality violation of Ref.~\cite{orsay17}. Moreover, their analysis
clearly supersedes the discussions of Isgur \cite{isgur16}, and that
chapter can be considered closed. Yet we note a question
raised in the footnote about the validity of some conclusions of
Ref.~\cite{d2}. According to the above illustrations, such
qualifications should be regarded as groundless. 
\vspace*{2mm}

A general note can be made here regarding
the attempts to challenge the applicability of the OPE. There are two
basic ingredients involved in the `practical OPE'. One is constructing
the large-$\omega$ expansion of the transition amplitudes like ${\cal
A}(\omega)$ in Eq.~(\ref{52}) in terms of the local heavy quark
operators. The second step is relating this asymptotic expansion to
the actual width. It is understandable that the challenge is motivated
by the (sometimes indeed not very intuitive) fact that the final expression
for the total width is given by the expectation values of the local
heavy quark operators. However, it turns out that all the concrete
objections in the literature or those hanging around as folklore attempted
{\it de facto} to undermine the second step, which is a purely
mathematical procedure and is indisputable. 

It is the first step which is far less trivial, both physically and
technically. Namely, here we relate the high-energy asymptotics of the
transition amplitude in QCD to the short-distance expansion of the
heavy quark Green function in the external field averaged over the
nonperturbative QCD ensemble of quarks and gluons (recall Fig.~1). 
And, curiously enough,
just this least trivial part has not been challenged in the
publications attempting to disprove the theory of the $1/m_Q$
expansion. 

%%%%%%%%%%%%
\subsection{Local duality proper}
%%%%%%%%%%%%

We do not have to say much regarding the literature in this respect, 
for -- to our
knowledge -- motivated estimates of the significant effects in
total semileptonic widths have never been 
presented.\footnote{The situation at times contains an ironic twist
when habitual critics of local duality in the integrated semileptonic
widths present quite precise relations of their own on {\it partial}
spectra in semileptonic or radiative $B$ decays -- with such concerns
becoming muted. For instance, it has been put forward (see,
e.g. \cite{neubnew}) that a measurement of part of the photon spectrum
in $B\to X_s+\gamma$ enables 
us to predict the $B\to X_u\,\ell\nu$ rate restricted to a
small and most vulnerable slice near the end point with a few 
percent accuracy. One has to keep in mind here that such a relation 
\cite{motion} rests solely on the OPE and actually holds 
only to the leading order
in $1/m_Q$. It is not otherwise supported by heavy quark
symmetry or any other independent argument.

Similarly, reservations regarding the accuracy of duality for
total semileptonic widths can be found 
in Ref.~\cite{ligeti}, while estimating in Ref.~\cite{lig} 
the uncertainty in the $b\to u$ rate over the very limited 
(and marginally hard) domain of maximal $q^2$ to be only a few
percent, including the effects of local duality.}
The realistic estimates we mentioned all yield tiny duality
violation. The possible rationale has been illustrated in 
Sect.~6.2.

\section{Quark-hadron duality and extracting  $|V_{cb}|$}

One of the central phenomenological applications of the heavy
quark expansion in beauty decays is extracting the value of the CKM
parameter $V_{cb}$. Two quantities are believed to be best suited 
for this purpose yielding the highest precision with limited model
dependence. One is the total semileptonic width 
$\Gamma_{\rm sl}(b\to c)$; another is the exclusive rate for 
$\Gamma(B\to D^*\ell\nu)$ extrapolated to zero recoil, where the $B\to
D^*$ formfactor $F(0)$ is fixed if {\em both} $m_b$ and $m_c$ are sufficiently 
heavy.\footnote{Both methods were discussed
and even applied already in the 80's. Their actual value has been
revealed with developing the theory quantifying the preasymptotic
corrections \cite{rev}.}

The experimental advantages are well known: the total semileptonic $b\to
c$ rate is by now one of the best measured quantities in $B$ physics. 
Measurement of the $B\to D^* \ell\nu$ rate is far more
involved, but due to the pioneering work by ARGUS and CLEO,
and also with LEP and recent BELLE data has been determined with 
decent accuracy. Yet the necessary extrapolation to the point of zero recoil
still introduces sizeable uncertainties.\footnote{The claims of a
reliable theoretical prediction of the shape of the $B\to D^*$
formfactors quoted, e.g. in Ref.~\cite{babar}, essentially overstate
their alleged model-independence.} The recent significant jump in
the CLEO value of the zero-recoil decay amplitude is an illustration
of these problems. 

Here we address the purely theoretical aspects of the
comparison of the two methods to sort out  
facts from fiction concerning these approaches (a brief review of the
historical misconceptions can be found in Ref.~\cite{rev}). In particular, the
question of `quark-hadron duality' has been advertised as an 
{\tt additional assumption} inherent and specific to 
$\Gamma_{\rm sl}(B)$. 

Once again, we have to stress that the OPE expresses 
the width  as a power series in the inverse energy release
\beq
\frac{\Gamma_{\rm sl}(B)}{\Gamma_0} = a_0 + a_2\frac{1}{E_r^2}
 + a_3\frac{1}{E_r^3} + \ldots \;.
\label{150}
\eeq
All such series are asymptotic and do not define the sum with 
unlimited accuracy at finite masses. As pointed out by Shifman
\cite{shifdual}, it is this feature of the expansions that underlies
violations of {\it local duality}. Clearly the same is true for the 
$B\to D^*$ formfactor at zero recoil:
\beq
F(0) = b_0 + b_2\frac{1}{m_c^2}
 + b_3\frac{1}{m_c^3} + \ldots \; , 
\label{152}
\eeq
where we have dropped the numerically less significant $1/m_b$ 
corrections. 
This series is likewise asymptotic -- the expansion Eq.~(\ref{152})
does not define completely $F(0)$ as a function of quark masses. 
The difference in respect to general quark-hadron duality between the
two cases, therefore lies in the details of the expansions potentially
affecting the quantitative behavior. 

The two series of Eqs. (\ref{150}) and (\ref{152}) superficially look 
quite similar; most notably, neither contains a correction 
linear in $1/m_Q$. Yet there are substantial differences. 
The expansion for $\Gamma_{\rm sl}(B)$, Eq.(\ref{150}), is given by 
$B$ meson expectation values of {\it local heavy quark
operators}. From a theoretical viewpoint,  
they are of a rather universal nature, and their size can be related 
to other
short-distance observables. In the series for $F(0)$, Eq.(\ref{152}), on
the other hand, {\em none} of the power corrections is related to  
local operators. Instead they are expressed by typical 
long-distance correlators shaped by dynamics of momentum scales 
$\sim {\cal O}(\Lam)$. Even the leading power terms are basically 
unknown. It should be noted that the emergence of any such term 
in the series for $\Gamma_{\rm sl}(B)$ would constitute a breakdown of 
the OPE and invalidate the whole theory of the inclusive 
widths. 
  
As stated above, duality violation is related to the effects of
sufficiently high orders which for asymptotic expansions may not fade
out quickly enough;   
including higher orders will not improve 
the accuracy of the result. While this represents a legitimate 
question for Eq.(\ref{150}), the question itself is somewhat 
ambiguous for $F(0)$ in 
Eq.(\ref{152}), since there already the first terms are 
unknown in practice. 

Nevertheless let us ignore this practical problem and just 
assume that all terms in Eq.(\ref{152}) can be 
determined with unlimited accuracy.  Then we can address 
the issue of duality for $F(0)$ and compare it with the 
situation in $\Gamma_{\rm sl}(B)$.

According to conventional wisdom the series (\ref{152}) 
approximates the exact amplitude with exponential accuracy 
for nearly Euclidean quantities like $F(0)$:
\beq
\delta_D F(0) = F(0) - F(0)_{1/m_c\,{\rm expansion}} \propto
{\rm e}^{-m_c/M_h}\;,
\label{154}
\eeq
with $M_h$ representing the characteristic hadronic scale (for clarity
we keep only the leading corrections related to $m_c$; i.e., 
we put $m_b
\to \infty$ while keeping $m_c$ finite). 

Such an exponential suppression of the deviation is not necessarily the case 
for a  
Minkowskian OPE observable like $\Gamma_{\rm sl}(B)$. Even though the
appropriately averaged widths coincide with their expansion
exponentially in $E_r$, the point-to-point oscillating difference can  
in principle be only power suppressed:
\beq
\frac{\Gamma_{\rm sl}(B)}{\Gamma_{\rm OPE}} -1 
= \frac{\varphi(E_r)}{E_r^p}\;, \qquad
\aver{\varphi(E_r)} = 0
\;.
\label{156}
\eeq
Thus, the assumption of {\it local quark-hadron duality} is the
approximation that one can discard, without averaging but at a given 
large value of the $b$
quark mass and $E_r$ the power-suppressed r.h.s.\ of Eq.~(\ref{156})
like the ``usual'' exponential terms in the  r.h.s.\ of Eq.~(\ref{154}). 

Because of the different functional behavior we naturally expect the OPE
to apply better numerically in the Euclidean than in the Minkowskian 
domain: there may be the oscillating component in inclusive
probabilities. However, this general observation obviously requires a
crucial qualification: the observables and, in particular, their
{\it ``hardness'' in energy scale} must be commensurate. 

In practice, however, the situation is rather different in
$F(0)$ vs. $\Gamma_{\rm sl}(B)$. The latter is much better for the OPE
than a generic inclusive probability, say than $R_\tau$ (not to
mention $R_{e^+e^-}$ or $\Gamma_{\rm nl}(B)$). Even more importantly, the
large mass expansion parameter is far better in $\Gamma_{\rm sl}(B)$
($E_r\!\simeq \!3.5\GeV$) than in $F(0)$ ($m_c \!\simeq\!
1.25\GeV$). This underlies the actual hierarchy. While at large enough
$m_c$ the exclusive $F(0)$ would evidently enjoy a smaller {\tt duality} 
uncertainty, in
practice $\Gamma_{\rm sl}(B)$ is much more robust in this respect. 

The exponential suppression (\ref{154}) in
the Euclidean case will actually set in only for masses that sufficiently 
exceed the scale of the strong interactions. Duality violation in
the semileptonic width is numerically too strongly suppressed in this
case to be relevant
in practice regardless of being exponential or only powerlike. We  
illustrate this point in numbers. 

Let us recall that the leading long-distance term in $F(0)$,
Eq.~(\ref{152}) constitutes an about $-7\%$ effect \cite{vcb}; 
i.e., 
the correction in the decay rate is $-15\%$ (cf.\  $-5\%$ in
$\Gamma_{\rm sl}(B)$). What can be the magnitude of the exponential
terms Eq.~(\ref{154}) which are not captured in the straightforward  
$1/m_Q$ expansion? With $m_c\!=\!1.25\GeV$ the size of 
$\delta_D F(0)$ would constitute
dozens percent even for the moderate value $M=
600\:\mbox{to}\:800\MeV$. However, we can argue that the actual high
momentum scale set up by quark masses is here $2m_c$ rather than 
$m_c$ itself.\footnote{This
is related to the position of the other singularities in the
transition amplitude. This feature is explicit in the concrete 
lowest order calculations.} Then one has
\beq
\delta_{D} F(0) \propto {\rm e}^{-2m_c/\mu_{\rm hadr}} \simeq
\left\{
\begin{array}{ll}
1.5\% & \mu_{\rm hadr}=0.6\GeV \\
4\% & \mu_{\rm hadr}=0.8\GeV \\
8\% & \mu_{\rm hadr}=1\GeV 
\end{array}
\right.
\label{160}
\eeq
While this simple estimate is only tentative, it clearly shows that
the $1/m_c$ expansion has intrinsic uncertainties here at least at a
percent level even in the most optimistic scenario. This is in accord
with the realistic estimates of the magnitude of the second and third 
order effects, $\delta_{1/m^2} F(0)\simeq -7\%$ \cite{buv}, 
$|\delta_{1/m^3}F(0)|\gsim 2\%$ \cite{varenna,ioffe}.\footnote{The
idea of insignificant higher order contributions to $F(0)$ has its
roots in the original estimates of this formfactor by Neubert being
very close to unity: $\delta_{1/m^2}F(0)= (-2\pm 1)\%$
\cite{neubf0}. Such scale of the leading nonperturbative
effects would indeed suggest negligible higher-order 
and exponential terms.}  

The estimates of $\delta_D F(0)$ can be confronted with the
local duality violation in the semileptonic $b\to c$ decay width. At
the total energy amounting to the  charm mass $1.25\GeV$ it may
naturally  
be large. However, the key fact -- the significant energy
release $E_r\simeq 3.5\GeV \gg m_c$ makes things completely
different. All estimates given in the preceding sections as well as in
other dedicated analyses, yielded the magnitude {\it much below} a
percent level. 

It should be noted that moderate uncertainties in $F(0)$ are 
obtained only in optimistic scenarios where $\mu_{\rm hadr}$ does 
not exceed $0.7\GeV$. If that indeed were to reflect the true   
scale of nonperturbative dynamics, then violations of {\em local} 
duality would simply be invisible in $\Gamma_{\rm sl}(B)$. 

As first explained in Ref.~\cite{rev}, the scenario with $\mu_{\rm
hadr} \simeq 1\GeV$ or even somewhat larger is not ruled out so far 
phenomenologically, even though there is no compelling direct evidence 
in its favor. To find possible significant duality violation in
$\Gamma_{\rm sl}$, we allow the relevant $\mu_{\rm hadr}$ to be as
large as $2$ to $2.5\GeV$. The quark-hadron duality `violation'
$\delta_D F(0)$ does not need to blow up literally in this
case, of course: even in the absence of a real exponential (in
$m_c$) factor it may be suppressed by numerical factors of order
$1$. For instance, at $m_c \lsim \Lam$ the corresponding zero recoil
formfactor for $B\to K^*$ is expected to be
$0.5\:\mbox{to}\:0.6$. Even in the worst scenario one still
might have $F(0)$, say around $0.8$. The point is that in such a case
the estimate of $F(0)$ in the $1/m_Q$ expansion would be not much more
sensible than trying to obtain $F_{B\to K^*}$ in a $1/m_s$ expansion. 

This discussion shows that contrary to popular lore 
total semileptonic widths suffer considerably less from duality 
violations than the zero-recoil $B\to D^*$ rate for the actual 
values of $m_b$ and $m_c$. The violation of {\it local duality}
inherent to decay widths, is strongly suppressed by the
peculiarities of the total semileptonic width, and the numerical
hierarchy is completely reversed in favor of $\Gamma_{\rm sl}(B)$ by the 
significant energy release in $b$ decays, 
which is much higher than the
charm quark mass. Some damping exponents are still larger that other
``slow powers''.

%%%%%%%%%%%%%%%%%%%
\section{Conclusion and Outlook}
%%%%%%%%%%%%%%%%

In this paper we have critically reviewed 
the salient aspects of treating nonperturbative QCD relevant to
quark-hadron duality, and in particular to violation of local
duality as it applies to the total semileptonic widths of heavy flavors.
A new underlying mechanism  was addressed which has been suggested
previously to be potentially significant, but not analyzed before.
Still we have found in 
Sect.~6 that conservatively it cannot exceed a fraction of
percent level in $\Gamma_{\rm sl}(B)$ 
with realistic estimates yielding even much smaller numbers. 
Thus, the corresponding uncertainty is totally negligible
in practical extraction of $V_{cb}$. 

It should be clearly realized that this conclusion applies only to the
{\it totally integrated} semileptonic widths. Most of the suppression
factors we discussed do not apply even qualitatively to partially
integrated widths, in particular restricted to the limited regions of
the lepton phase space. This reservation is supported by the
explicit model estimates. For instance, in the 't~Hooft model the
local duality violation in the differential distributions, in
particular, in $b\to s+\gamma$ or $b\to u\,\ell\nu$ was found to be very
significant numerically \cite{lebur} for practically relevant kinematics.

Typically, we expect duality violation to affect $\Gamma_{\rm sl}(B)$
only at a permill level. However, this
does not mean we can compute in practice 
$\Gamma_{\rm sl}(B\!\to \!X_c)$ today with
such an accuracy. For we specifically addressed the {\it
violation of local duality} proper here, and considered only the
related uncertainties. The theoretical progress in
the recent years allowed us to formulate the question consistently;
this is far from trivial and has often been missed in the
literature. In simple terms, this uncertainty tells us how accurately 
we can determine the
width if as many as necessary terms in the (practical) OPE are known 
with sufficient precision.

{}From the present and preceding
theoretical studies of quark-hadron duality and its violation
we can draw a final practical conclusion.  
The accuracy with which $|V_{cb}|$ can be extracted from $\Gamma_{\rm sl}(B)$
(and $|V_{ub}|$ from $\Gamma_{\rm sl}(B\!\to \!X_u)$) is actually limited 
only by the precision with which we know the first few 
terms in the OPE. Duality violations are negligible in practice. 
Even approximate knowledge of the magnitude of higher order 
nonperturbative contributions could be of help in a complementary
respect: to determine the mass scale $\mu_{\rm hadr}$ which provides 
the effective yardstick for nonperturbative dynamics in heavy quarks.
$\mu_{\rm hadr} < 1\GeV$ would imply a ``favorable scenario'' where 
one can hope to apply heavy quark expansion to quantify deviations
from the heavy quark symmetry in charm. If it were to exceed $1\GeV$, 
one would probably have to abandon such methods for model-independent
determination of underlying weak decay parameters in general, and 
rely only on the more robust cases of the
inclusive observables in beauty decays. Even such a scenario would not 
elevate local 
duality violation in $\Gamma_{\rm sl}(B)$ to practically significant
levels. Yet it would affect extracting  the 
fundamental input parameters like quark masses and the first
nonperturbative parameters $\mu_\pi^2$, $\mu_G^2$, $\rho_D^3$ from the
data. As we point out below, there is a direct experimental way to
infer the scale of $\mu_{\rm hadr}$ in $B$ decays.

All considered mechanisms yield quite small violations of local
duality in $\Gamma_{\rm sl}(B\!\to\! X_c)$ when 
compared to the impact on $R_\tau$ from 
$\tau$ decays, which has become the canonical
yardstick for judging the OPE. This had actually been anticipated. The
qualitative arguments can be found in Ref.~\cite{varenna}; now we can
make the comparison in a more elaborated manner illustrating this 
in quite general terms.

$\bullet$
The one argument given in the literature to substantiate smallness
of local duality violation in $R_\tau$ invokes a strong threshold
suppression,\footnote{We remark here that -- contrary to common lore -- 
increasing this power in reality deteriorates duality: 
for a large power $n$ 
$\:R_\tau$ is saturated at $s\lsim M_\tau^2/n$, and eventually 
it simply ceases to be a short-distance quantity. Suppression of
local duality violation with increase of
$n$ would only apply in the academic case where $M_\tau^2$ scales with
$n$ so that $M_\tau^2/n$ remains much larger than $\mu_{\rm
hadr}^2$. In practice half of $R_\tau$ comes from $s$ in
Eq.~(\ref{40}) below $1\GeV$.} $\propto \Delta^2$:   
the hadronic states
$X$ with $M_X \!=\! M_\tau\!-\!\Delta$ contribute to $R_\tau$ with the
weight $\left(\frac{\Delta}{M_\tau}\right)^2$. Yet the threshold
suppression is {\em much stronger} in $\Gamma_{\rm sl}$ where it is 
the {\tt fifth} power, $\Delta^5$. 

$\bullet$
When decreasing the energy scale $R_\tau$ blows up and
duality is $100\%$ violated. $\Gamma_{\rm sl}(B\!\to\! X_c)$, on the
contrary, exhibits very accurate local duality even with shrinking
energy release; this is ensured by the SV limit.

$\bullet$
Local duality violation in $\Gamma_{\rm sl}(B\!\to\! X_c)$ comes
from highly excited charm states with $M_{X_c} \!\approx \!M_B$; they are thus
driven by perturbative gluon exchanges. In their
absence the yield in the relevant domain -- 
denoted by $P$ in Fig.~3 -- would practically vanish.
The relative scale of such an effect starts at the  $\alpha_s/\pi \lsim
15\%$ level, even before any other suppression factors are considered. 
In contrast, duality in $R_\tau$ or
$R(s)$ affects already the ``valence'' quark contribution which is
roughly a constant $N_c$ at arbitrary energy. 

$\bullet$
The energy scale itself is obviously much higher in
$\Gamma_{\rm sl}(B\!\to \!X_c)$  than in $\tau$ decays. \vspace*{1.5mm}

With all these effects acting in the same direction, one ends up
with a very small local duality violation in $\Gamma_{\rm sl}(B\!\to\!
X_c)$ even allowing for it to be as large as about $5\%$ in
$R_\tau$. In view of this comparison, discussing possible significant
duality violations in the semileptonic beauty widths and not allowing
for appreciable effects in $\tau$ decays does not seem to constitute a
consistent application of QCD.

The suspicions towards large effects of local duality violations in
the decay widths were traditionally fed by reported problems in
nonleptonic widths, possibly showing up in the size of 
${\rm BR}_{\rm sl}(B)$ and of the $\tau_{\Lambda_b}/\tau_{B^0}$ ratio
\cite{ioffe}. 
However, the maximal magnitude of local duality 
violations in
nonleptonic decays is not limited, in principle, by any of the 
features peculiar to semileptonic widths as 
discussed in Sects.~4 and 6, which tend to reduce duality
violation there by orders of magnitude.

Another potential problem discussed is the absolute semileptonic decay
rate of $D$ mesons: about a third of the actual width may seem to be
unaccounted for in the $1/m_c$ expansion; this deficit 
could be interpreted as due to duality violation. However, as pointed out in
Ref.~\cite{vub}, the excess of the 
observed decay rate can naturally 
be explained by usually discarded nonvalence four-quark expectation 
values; they would still 
yield a marginally noticeable correction in total $B$ decay widths. 
This conjecture was reiterated in Ref.~\cite{lebur}. In any case, it is
natural to expect sizeable, or even large duality
violation in charm while having it very small in beauty.
\vspace*{2mm}

A rather direct experimental check of local quark-hadron duality
and its onset in a broad enough range of energies would be a
measurement of the decay rate distribution over the
invariant final state hadron mass $M_X$ 
(or, equivalently, of a combination of the
structure functions $w_1$ and $w_2$) in the $B\to X_c\,\ell\nu$ decays
in the mass region above $3\GeV$.\footnote{The resonance region below it
essentially determines the OPE parameters $\La(\mu)$,
$\mu_\pi^2$, $\mu_G^2$, $\rho_D^3$ etc.} This would provide unique
and detailed information on QCD in the transition from the
nonperturbative to the perturbative regime, at the level of radiative
effects directly in Minkowski space. The primary goal here would be to
analyze the onset of the perturbative regime, domain $P$ in Fig.~3, by
comparing the actual hadronic yield with the perturbative result. (The
structure function $w_3$ becomes accessible if the electron energy is
additionally measured.)

The total semileptonic width corresponds to integrating these
structure functions over the whole perturbative domain and also
dilutes them by adding the dominant contribution from the resonance
region $R$. This makes duality violation invisible in $\Gamma_{\rm
sl}(B)$. However, local duality violations are expected to be evident in
the differential distribution, and it should be checked that the
structure functions oscillate around their perturbative expressions,
in particular in the lower part of the interval in $M_X$ -- but
coincide with the latter upon averaging. As a byproduct of such an
analysis one can obtain an independent {\it direct} measurement of the
effective QCD coupling $\alpha_s$ at low energies, and evaluate the
magnitude of the higher-order local heavy quark operators entering
heavy quark expansion. 

A similar -- although, probably, more remote experimentally --
possibility is to study in detail the similar $M_X$ 
distribution in $b\to u$ transitions -- much like it was proposed for
model-insensitive extraction of $|V_{ub}|$ \cite{mx}. Here even a more
detailed information on the duality
onset is accessible in principle {\it via} genuine double or 
triple distributions. In practice, though this may be obscured
by the necessity to restrict kinematics to exclude $B\to
X_c\,\ell\nu $ decays and by intervention of the
standard nonperturbative OPE corrections (in particular, due to Fermi
motion) up to $M_{X_u}=1\:\mbox{to}\:1.5\GeV$.

In general, we think that all practical methods of extracting
$V_{ub}$ suggested so far would bear a more or less significant element
of model dependence until local duality is carefully studied in $B\to
X_c\,\ell\nu$ along the lines suggested above.
\vspace*{2mm}

While the central role of the OPE is becoming widely accepted now, certain 
subtle, yet significant features of its implementation are often 
not fully appreciated despite extensive reviews and successful
applications in the literature. 
It is often overlooked that only a careful treatment of the OPE
yields  accurate and defendable results:  

{\footnotesize  $\clubsuit$}  
One should use properly defined
cutoff-dependent strong interaction parameters: quark
masses $m_{b,c}(\mu)$, kinetic $\mu_\pi^2(\mu)$ and chromomagnetic 
$\mu_G^2(\mu)$
expectation values, the Darwin term
$\rho_D^3(\mu)$ etc. Ill-defined and indefinite pole masses, $\lambda_1$ of
HQET etc.\ should be avoided. 

{\footnotesize $\clubsuit$} 
In this approach there are no significant
uncertainties in the radiative corrections or those associated with
the value of the $b$ quark mass. This has been substantiated by 
explicit computations of the full ${\cal O}(\alpha_s^2)$ corrections
\cite{czarmelntimo}.

{\footnotesize $\clubsuit$} There is limited direct impact from 
the uncertainty in
$\mu_\pi^2$ and $\mu_G^2$, as well as from the higher-order
terms.\vspace*{1.5mm}

The analysis shows that the current limiting factor in determining 
$V_{cb}$ is the {\tt precise value of} $\,m_c$. At present the most
accurate way is relating $m_c$ to $m_b$ using the expansion of
hadron masses \cite{optical}:
\beq
m_b\!-\!m_c \!=\!\frac{M_B\!+\!3M_{B^*}\!}{4}-
\frac{\!M_D\!+\!3M_{D^*}\!}{4} +
\frac{\mu_\pi^2}{2} \!\left(\frac{1}{m_c\!}\!-\!\frac{1}{m_b}\right) \;+
\frac{\rho_D^3\!-\!\bar\rho^3\!}{4}\!
\left(\frac{1}{m_c^2\!}\!-\!\frac{1}{m_b^2\!}\right) + {\cal
O}\!\left(\!\frac{1}{\,m^3\,\!}\!\right).
\label{170}
\eeq
This is the only place in evaluating the semileptonic width where 
we rely
on the expansion in $1/m_c$ and thus are sensitive to
the values of the higher condensates. Theoretically, the  
nonrelativistic expansion of mass is known to behave the best
compared to, say, the expansion of wavefunctions and their overlaps
determining, for instance the $B\to D^*$ formfactor. There exist a
number of independent determinations of the short-distance charm mass
$m_c$, see, e.g.\ Refs.~\cite{mc} 
for $m_c(m_c)$. Translated to the running masses
used in the heavy quark expansion, they yield very similar numbers 
using just the central values of $m_b$, $\mu_\pi^2$ and the expected
moderate values of the higher-order condensates, without even invoking
the error bars.

Recently a direct determination of the heavy quark mass and the kinetic
expectation value has been reported by CLEO
\cite{cleomom}. Unfortunately, the preliminary version gives the 
fitted parameters
$\La$ corresponding to the pole mass, and $-\lambda_1$ 
which makes the results generally
unusable. However, using the quoted values of the hadronic moments
themselves and the estimate of the Darwin term $\rho_D^3=0.12\GeV^3$
\cite{four} 
one obtains central values for $m_b(\mu)$ and $\mu_\pi^2(\mu)$ in
a good agreement with the theoretical expectations \cite{ioffe}:
$\La(1\GeV) \simeq 670\MeV$, $\mu_\pi^2(1\GeV) \simeq
0.42\GeV^2$. We note parenthetically that such an analysis, as a
matter of fact, is most sensitive to the value of $m_c$ (more
precisely, to $m_b\!-\!m_c$). In this respect, it is largely suited
to determine the possible size of the higher-order terms in the mass
expansion Eq.~(\ref{170}). \vspace*{2mm}

{\it To conclude:} future improvements in the theoretical accuracy
in $V_{cb}$ are expected to occur through better control over the
higher-order terms in the meson mass expansion Eq.~(\ref{170}) and/or an
independent precise determination of the low-scale running {\tt charm}
mass.

%%%%%%%%%%
\section{Epilogue}
%%%%%%%%%%%

As a final note let us express how we view as 
quite paradoxical the way in which the discussion of 
quark-hadron duality and its 
limitations in $B$ decays has -- and has not -- taken place. 

On one hand there is the exclusive transition $B \to D^*\, \ell\nu$: 
its formfactors $F_{D^*}$ cannot be expressed through 
an expansion of local operators even at zero recoil and the  
leading power corrections $\sim {\cal O}(1/m_c^2)$ are 
not fully known; they are estimated relying on natural 
assumptions \cite{vcb,rev}
which, however become vulnerable when descending to the level of a few percent 
error bars. Nevertheless statements of an unrealistically small 
theoretical uncertainty are typically accepted
as gospel without much reflection; it is forgotten  
how much the central value stated for $F(0)$ has changed over 
the years. 

The psychological background behind the apparent tolerance is quite 
understandable: Already the
leading power corrections $\propto 1/m_c^2$ in $F(0)$ are
unknown and the estimates of the small-uncertainty variety come 
from {\it ad hoc} model assumptions. There is no much to criticize 
here beyond that.

Yet for the fully inclusive semileptonic width $\Gamma_{\rm sl}(B)$, 
for which the consistent OPE has been given, suggestions of
uncontrollable theoretical errors are readily picked up -- despite  
the fact that dedicated theoretical analyses have 
given no reproducible sign of such effects and that the theoretical 
predictions have not changed in any significant way over the years. 
Part of the reason might well be that even the central result -- the 
absence of ${\cal O}(1/m_Q)$ corrections in the width \cite{buv,bs} -- 
is highly nontrivial and becomes intuitive only within the proper approach.
The OPE itself, while well developed, remains a conceptually nontrivial
theoretical technology employing a number of basic tools. This 
provokes, in our opinion, much critical attention to the total
widths and attempts to challenge the heavy quark expansion for
$\Gamma_{\rm sl}(B\!\to\! X_c)$ at a more pedestrian level.

\vspace*{3mm}

\noindent
{\bf Acknowledgements:}
This %% work 
Vademecum was to a large extent initiated by our exchanges with 
experimental colleagues, in particular from the CLEO collaboration and the
LEP Heavy Flavour Working Group. N.U is pleased to acknowledge
useful discussions with A.~Le~Yaouanc, L.~Oliver,
O.~P\`ene and  J.-C.~Raynal on related subjects. 
We are deeply grateful to M.\,Shifman for critical comments.
I.B.\ thanks the INFN Group in Milano
for their hospitality during completing this paper.  
This work was supported in part by the NSF under grant number PHY-0087419.

%%%%%%%%%%%%%
\section{Appendix}
%%%%%%%%%%%

The analysis briefly described in this Appendix is quite general in
nature. Therefore, we do not specify explicitly the type of observable
we discuss. Likewise, we generically denote the large energy scale
parameter as $E$ (it is $m_Q$ or $E_r$ in the heavy quark decays;
likewise, it can actually be $m_Q^2$ or $s=E^2$);
however, to keep in the context with our discussion we will refer to the
short-distance observable as $\Gamma_B$. 

Under our convention explained in Sect.~6.2, we discard the
higher-order terms in the OPE, and 
\beq
\frac{\Gamma_B-\Gamma_{\rm OPE}}{\Gamma_0} \propto 
\frac{1}{E^k} f(E\!-\!n\Delta)
\label{99}
\eeq
and $\Gamma_{\rm OPE}/\Gamma_0= b_0+b_1/E+...+b_k/E^k$ (the coefficients
$b_k$ can logarithmically depend on $E$). Likewise we can
discard all effects in $\Gamma_B$ which fade out faster than $1/E^k$,
since they are subdominant in $\chi(E)$. The requirements stated in
Sect.~6.2 are then sufficient to determine the exact asymptotic form of
the local duality violation, i.e.\ the function $f(E)$, including its
overall normalization.

Indeed, let us consider $(\Gamma_B\!-\!\Gamma_{\rm OPE})/\Gamma_0$ on the
interval between the opening of the two successive thresholds. The
contribution of the last open channel in  $\Gamma_B/\Gamma_0$ is given
by 
$$
\frac{c}{E^k} (E\!-\!n\Delta)^\gamma \theta(E\!-\!n\Delta) 
\left(1+{\cal O}\left(\frac{\Delta}{E}\right)\right)\;.
$$
We do not need to know explicitly either contributions of the other,
lower mass channels (they are expandable in $1/E$), or $\Gamma_{\rm
OPE}$ -- it suffices to know that their contribution to $\Gamma_{\rm
OPE}/\Gamma_0$ is given by a few powers of $1/E$. Expanding the
difference of these terms around $E_n=n\Delta$ we get a general
polynomial
\beq
\frac{c}{E^k}\left(a_0+a_1(E\!-\!E_n)+ a_2(E\!-\!E_n)^2 + \ldots +
a_{k}(E\!-\!E_n)^{k}
\right)\;.
\label{100}
\eeq
The duality violation is then given, up to a factor $c/E^k$, by the
difference
\beq
g_\gamma\left(\frac{E}{\Delta}\right)=
(\Delta E)^\gamma-a_0-a_1\Delta E- a_2(\Delta E)^2 - \ldots -
a_{k}(\Delta E)^{k}
\label{101}
\eeq
$$
\Delta E \equiv \Delta \,{\rm Frac}\left(\frac{E}{\Delta}\right)\;,
\qquad\qquad \chi(E)= c \,\frac{g_\gamma(\frac{\Delta E}{\Delta})}{E^k}\
$$
(${\rm Frac}$ and ${\rm Int}$ denote the fractional and integer
parts, respectively). 

The function $g_\gamma\left(\frac{\Delta E}{\Delta}\right)$ is constrained by
the requirements that\\
\hspace*{1em}$\bullet$ its average vanishes\\
\hspace*{1em}$\bullet$ it is continuous, as well as its derivatives up to the
order $l\!=\!{\rm Int}(\gamma)$. Higher derivatives have discontinuity at
the threshold values $E\!=\!n\Delta$, determined by the threshold behavior
$(\Delta E)^\gamma$. \\
These impose $l+2$ linear constraints on the coefficients
$a_i$. Therefore, the consistency demands $k\!\ge\! l+1$. 
In the cases we
consider one actually has $k\!=\!l+1$. This is not always the case; the
thresholds can appear at a level suppressed by a higher power $k$ than
the threshold exponent $\gamma$. However, it can be argued that in
this case the last few (viz., ${\rm Int}(k\!-\!\gamma)\!-\!1$) terms in the
polynomial are absent, since it emerges from expanding the terms power
suppressed at least as $E^{-{\rm Int}(k\!-\gamma-1)}$. This
observation is quite general and is related to a certain property of
the OPE which is similar to {\it global duality} \cite{optical}. We do
not pursue this aspect here, and simply make use of the fact that $k\!=\!
l+1$ in our applications. Then the whole function describing the
asymptotic violation of local duality to the leading order in $1/E$ is
fixed in terms of the large-$E$ threshold behavior. 

A few relevant examples are given below. First, however, we note that
the branching fraction of the last open channel is $c\frac{(\Delta
E)^\gamma}{E^k}$, and it is maximal just at the threshold for the next
resonance. Therefore, it is advantageous to normalize the amount of
duality violation to this maximal last resonance fraction 
${\rm BR}_{\rm last}$:
\beq
\tilde \chi(E)=\frac{\chi(E)}{ {\rm BR}_{\rm last}} =
g_\gamma\left(\frac{\Delta E}{\Delta}\right)\;.
\label{102}
\eeq
Likewise, the maximal magnitude of such a normalized amount of duality
violation $|\tilde \chi_0|$ is simply a universal number depending
only on the power of the threshold suppression.

Now, for the most singular threshold behavior possible in $3+1$ dimension
$\Delta^{\frac{1}{2}}$ we have 
\beq
g(x) = x^{\frac{1}{2}}-\frac{1}{6}-x\;,  \qquad\qquad 0\le x \le 1,  
\qquad\qquad\qquad |\tilde
\chi_0|=\frac{1}{6}
\label{104}
\eeq
For higher powers we get
\beq
\begin{array}{lll}
g(x)=\frac{x}{2} -\frac{1}{12}-\frac{x^2}{2} & 
|\tilde \chi_0|=\frac{1}{12} & \mbox{ at } \gamma=1\\

g(x)=x^{\frac{3}{2}}- \frac{1}{40} -\frac{x}{4}-\frac{3x^2}{4} & 
|\tilde \chi_0|\simeq 0.0282 & \mbox{ at } \gamma=\frac{3}{2}\\

g(x)=\frac{x^2}{2}-\frac{x}{6}-\frac{x^3}{3}
 & 
|\tilde \chi_0|\simeq 0.0160 & \mbox{ at } \gamma=2\\

g(x)=x^{\frac{5}{2}}+\frac{1}{168}-\frac{x}{16}-\frac{5x^2}{16}-
\frac{5x^3}{8}  & 
|\tilde \chi_0|\simeq 0.00948 & \mbox{ at } \gamma=\frac{5}{2}\\

g(x)=\frac{x^3}{2} +\frac{1}{120} -\frac{x^2}{4}-\frac{x^4}{4}
 & 
|\tilde \chi_0|= \frac{1}{120} & \mbox{ at } \gamma=3\\

g(x)=x^{\frac{7}{2}}+\frac{5}{1152}+\frac{x}{48}-\frac{7x^2}{64}-
\frac{35x^3}{96} -\frac{35x^4}{64} & 
|\tilde \chi_0|\simeq 0.00525 & \mbox{ at } \gamma=\frac{7}{2}\\

g(x)=\frac{x^4}{2}+\frac{x}{30}-\frac{x^3}{3}-\frac{x^5}{5}
 & 
|\tilde \chi_0|\simeq 0.00489 & \mbox{ at } \gamma=4\\

g(x)=x^{\frac{9}{2}}\!-\!\frac{3}{1408}\!+\!\frac{5x}{256}\!+\!
\frac{3x^2\!\!}{64} \!-\!
\frac{21x^3\!\!}{128} \!-\! \frac{105x^4\!\!}{256}  
\!-\!\frac{63x^5\!\!}{128}& 
|\tilde \chi_0|\simeq 0.00364 & \mbox{ at } \gamma=\frac{9}{2}\\

g(x)=\frac{x^5}{2}-\frac{1}{252}+\frac{x^2}{12}-\frac{5x^4}{12}-\frac{x^6}{6}
 & 
|\tilde \chi_0|=\frac{1}{252} & \mbox{ at } \gamma=5
\end{array}
\label{106}
\eeq
The case $\gamma=2$ is relevant for the $\tau$ width. It was 
obtained in Ref.~\cite{d2} by explicitly constructing the OPE. In
lower space dimensions a more singular behavior is possible. The case
$\gamma=0$ was considered in  Refs.~\cite{d22} for the 't~Hooft
model. 

\thispagestyle{plain}
\begin{figure}[hhh]
\vspace*{-.2mm}  
 \begin{center}
 \mbox{\epsfig{file=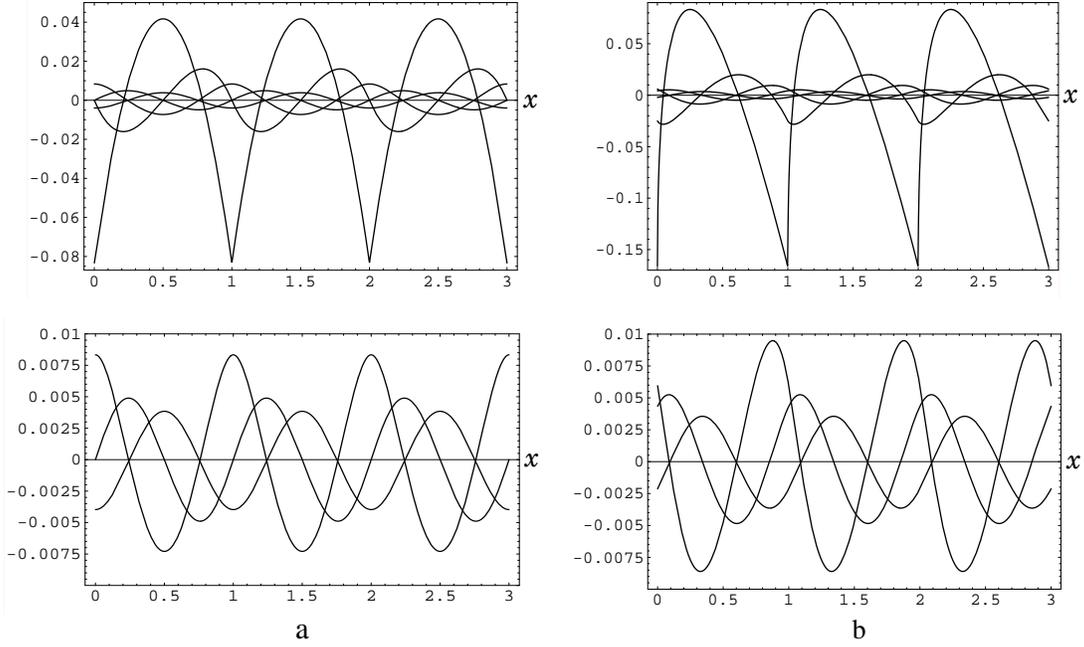,width=14.5cm}}
 \end{center}\vspace*{-5mm} 
\caption{ \footnotesize
Functions $g_\gamma(x)$ for integer $1\:\mbox{to}\:5$ ({\bf a})
and half-integer $\frac{1}{2}\:\mbox{to}\;\frac{9}{2}$ ({\bf
b}) threshold index $\gamma$. Curves with smaller magnitude correspond
to larger $\gamma$. Lower plots show magnified $g_\gamma(x)$ for
$\gamma=3,4,5$ and for $\gamma=\frac{5}{2},\frac{7}{2},\frac{9}{2}$,
respectively. 
}
\end{figure}

The corresponding functions are plotted in Fig.~4. The maximal
magnitude of the duality violation decreases fast with smoothening the
threshold behavior, which is transparent. Increasing $\gamma$ by one
requires vanishing of the average of one more derivative of the
function. This amounts to integrating $g(x)$ and subtracting its
average:
\beq
g_{\gamma+1}(x)= (\gamma+1)\left[\int_0^x g_\gamma (t) {\rm d}t + \int_0^1
t\,g_\gamma (t){\rm d}t \right]\;;
\label{108}
\eeq
its maximal deviation from zero drops fast. 
It is remarkable that the $g_\gamma(x)$ for $\gamma >1$ 
is very close (up to a factor) to $\cos{(2\pi x +\phi_\gamma)}$;
the origin is clear from the above recurrent relation. The latter 
simply states that the $n$-th Fourier mode of $g_\gamma(x)$ gets
multiplied by $i(\gamma+1)/(2\pi n)$; the higher modes then die out
quickly and only the leading one with $n=\pm 1$ survives. 
It is worth noting that the normalized deviations
$|\tilde \chi_0|$ do not depend even on the particular mass pattern --
the resonances can be equally spaced in mass or mass squared, which is an
added convenience.

It should be clear that the above evaluation of the duality violation
is asymptotic. At intermediate energies subleading in $1/E$ effects
can be noticeable; likewise the asymptotic behavior of the resonance
masses is distorted, which modifies the exact numbers.

\end{document}